\documentclass{article}
\usepackage{PRIMEarxiv}

\usepackage[utf8]{inputenc} % allow utf-8 input
\usepackage[T1]{fontenc}    % use 8-bit T1 fonts
\usepackage{hyperref}       % hyperlinks
\usepackage{url}            % simple URL typesetting
\usepackage{booktabs}       % professional-quality tables
\usepackage{amsfonts}       % blackboard math symbols
\usepackage{nicefrac}       % compact symbols for 1/2, etc.
\usepackage{microtype}      % microtypography
\usepackage{lipsum}
\usepackage{graphicx}

\usepackage{algorithm}
\usepackage{algpseudocode}

\usepackage{colortbl}
\usepackage[dvipsnames,table,xcdraw]{xcolor}

\definecolor{lightgray}{gray}{0.8} % adjust the shade as needed
\newcommand{\rev}[1]{\textcolor{black}{#1}}

\usepackage{longtable}
\usepackage{comment}

\usepackage{amsmath}
\usepackage{amssymb}
\usepackage{makecell}
\usepackage{multirow}
\usepackage{subcaption}
\usepackage{braket}

\newif\ifARXIV
\newif\ifIOP
\ARXIVtrue
\IOPfalse % IOP false

\title{QCPINN: Quantum-Classical Physics-Informed Neural Networks for Solving PDEs
}

\author{
  Afrah Farea$^{1}$, Saiful Khan$^{2}$, Mustafa Serdar Celebi$^{1}$\\
  \\[0.5em]
  $^{1}$Computational Science and Engineering Department, Informatics Institute \\
  Istanbul Technical University, Istanbul 34469, Turkiye \\
  \texttt{farea16@itu.edu.tr, mscelebi@itu.edu.tr} \\[0.5em]
  $^{2}$Scientific Computing, Rutherford Appleton Laboratory \\
  Science and Technology Facilities Council (STFC), OX11 0QX, United Kingdom\\
  \texttt{saiful.khan@stfc.ac.uk}  \\
}

\begin{document}
\maketitle

\begin{abstract}
Physics-informed neural networks (PINNs) have emerged as promising methods for solving partial differential equations (PDEs) by embedding physical laws within neural architectures. However, these classical approaches often require a large number of parameters to achieve reasonable accuracy, particularly for complex PDEs.
In this paper, we present a quantum-classical physics-informed neural network (QCPINN) that combines quantum and classical components, allowing us to solve PDEs with significantly fewer parameters while maintaining comparable accuracy and convergence to classical PINNs.
We systematically evaluated two quantum circuit architectures across various configurations on five benchmark PDEs to identify optimal QCPINN designs.
Our results demonstrate that the QCPINN achieves stable convergence and comparable accuracy, 
\rev{
while using only 10-30\% of the trainable parameters required by classical PINNs. This approach also results in a significant reduction in the relative $L_2$ error for Helmholtz, Klein-Gordon, and Convection-diffusion equations, with reduction ranging from 4\% to 64\% across various fields.
These findings demonstrate the potential of parameter efficiency and solution accuracy in physics-informed machine learning,  allowing for a substantial decrease in model complexity without compromising solution quality.} 
QCPINN presents a promising pathway to address the computational challenges associated with solving PDEs.
\end{abstract}

%---------------------------------------------------------------------------------------------------------%
\section{Introduction}
%---------------------------------------------------------------------------------------------------------%

Recent advancements in machine learning have transformed approaches to solving \rev{partial differential equations (PDEs)}, with \rev{Physics-informed neural networks (PINNs)} emerging as an innovative technique. PINNs embed physical laws directly into neural network architectures, eliminating the need for labeled training data while strictly enforcing physical constraints. This approach has demonstrated considerable success in modeling various physical phenomena across scientific and engineering domains~\cite{karniadakis2021physics,cai2021physics,farea2025learning,farea2025:HiPC}.
Although they hold great potential, research is underway to improve the accuracy of PINNs in modeling different PDEs and physical systems. 
They often encounter challenges with stiff systems, capturing steep gradients, and efficiently navigating high-dimensional parameter spaces \rev{for reasons such as memory requirements,  slow convergence, or when calculating high-order or high-dimensional derivatives~\cite{hu2024tackling,hu2025bias}}. 
While increasing model expressivity through additional parameters can alleviate these challenges, this approach introduces new problems: increased computational cost, risk of overfitting, and unstable convergence behavior during training~\cite{ying2019overview, farea2024:arXiv}. Such a trade-off between model complexity, generalization, and convergence ability is a fundamental challenge in neural network design. 

Quantum computing presents a promising paradigm for overcoming these limitations. With its ability to process high-dimensional data and exploit quantum principles such as superposition and entanglement,  quantum algorithms offer unique advantages for solving complex computational problems. For instance, the Harrow-Hassidim-Lloyd (HHL) algorithm has demonstrated theoretical exponential speedups for specific linear systems, while variational quantum algorithms provide practical near-term solutions~\cite{mitarai2018quantum, havlivcek2019supervised, cerezo2021variational}. Moreover, recent developments~\cite{knudsen2020solving,schuld2021effect,abbas2021power,du2021learnability} indicate that quantum neural networks possess an expressive power that could significantly enhance the learning capabilities for modeling complex physical systems.

Although both PINNs and quantum computing have advanced independently, their integration remains largely unexplored. Several preliminary efforts have emerged: Sedykh et al. proposed HQPINN~\cite{sedykh2024hybrid}, focusing on computational fluid dynamics in 3D Y-shaped mixers; Dehaghani et al. introduced QPINN~\cite{dehaghani2024hybrid}, integrating a dynamic quantum circuit with classical computing methodologies for optimal control problems; Trahan et al.~\cite{trahan2024quantum} investigated both purely quantum and hybrid PINNs. Leong et al. proposed HQPINN~\cite{leong2025hybrid}, which is inspired by Fourier neural operators. However, existing research has not comprehensively established whether quantum-enhanced models can improve upon purely classical PINNs across a broad range of PDEs efficiently and reliably.

We investigate the limitations of traditional PINNs by integrating quantum computing circuits with classical neural networks. We propose that such a hybrid approach can leverage complementary strengths: classical computing's robustness and quantum computing's enhanced expressivity with inherent parallelism. We believe this method will allow us to solve a wide range of PDEs with similar or better accuracy while requiring significantly fewer trainable parameters than classical models.

To this end, we introduce a hybrid architecture named Quantum-Classical Physics-Informed Neural Network (QCPINN). 
To search for a suitable QCPINN architecture and its configurations, we systematically evaluate multiple dimensions of the design space: two quantum circuit architectures:
(a) discrete-variable (DV) qubit-based circuits with configurations, such as four circuit topologies: Alternate, Cascade, Cross-mesh, and Layered; and 
two embedding schemes: Amplitude and Angle;
(b) continuous-variable (CV) circuits with configurations, including two measurement schemes: number and quadrature; 
three nonlinear operations: Kerr, cubic phase, and cross-Kerr; and 
two parameterizations: phase-free (magnitude) and full (magnitude and phase).
We evaluate the performance of these QCPINN models across various PDEs, such as the Helmholtz and lid-driven Cavity equations, while also extending our analysis to the Klein-Gordon, Wave, and Convection-diffusion equations. Our comprehensive assessment compares (a) solution accuracy, (b) convergence rates, (c) parameter efficiency \rev{(d) time, and (f) memory} between classical PINNs and QCPINN variants. 

To ensure reproducibility and facilitate collaborative advancement, we have made all components of QCPINN open-source code and publicly available on GitHub at~\url{https://github.com/afrah/QCPINN}~\cite{afrah2025qcpinnGithub}.

%---------------------------------------------------------------------------------------------------------%
\section{Related Work}
%---------------------------------------------------------------------------------------------------------%

In the domain of PDE solvers, quantum algorithms can be broadly categorized into two primary approaches: (a) exclusively quantum algorithms and (b) hybrid quantum-classical methods. 

\subsection{Exclusively Quantum Algorithms}

Exclusively quantum algorithms for solving PDEs typically leverage unitary operations and quantum state embedding techniques to achieve a computational advantage. In contrast to neural networks, exclusively quantum algorithms use fixed quantum circuit structures with no trainable parameters.
The celebrated Harrow-Hassidim-Lloyd (HHL) algorithm is an example of this category, which provides exponential speedups to solve linear systems under specific conditions~\cite{harrow2009quantum}. Extensions of the HHL algorithm have been applied to solve sparse linear systems derived from discretized linear~\cite{ambainis2012variable, somma2016quantum} or nonlinear~\cite{leyton2008quantum} PDEs, leveraging numerical techniques such as linear multistep method~\cite{berry2014high}, Chebyshev pseudospectral method~\cite{childs2020quantum}, and Taylor series approximations~\cite{berry2017quantum}. In fluid dynamics applications, lattice-gas quantum models have been examined for directly solving PDEs~\cite{yepez2002quantum,montanaro2016quantum} or PDEs derived from ODEs through methods such as the Quantum Amplitude Estimation Algorithm (QAEA)~\cite{oz2022solving,oz2023efficient}.

% Problem / Gap
Despite these advancements, purely quantum algorithms for solving PDEs encounter practical challenges such as requirements for quantum error correction~\cite{preskill1998reliable, terhal2015quantum, childs2021theory}, 
limitations in circuit depth~\cite{bravyi2005universal, montanaro2016quantum, boixo2018characterizing}, and bottlenecks in classical data embedding~\cite{aaronson2015read, ciliberto2018quantum, lloyd2013quantum}. These obstacles, among others, limit the near-term viability of exclusively quantum PDE solvers on today's noisy intermediate-scale quantum (NISQ) devices.

\subsection{Hybrid Quantum-Classical Approaches} 

Hybrid quantum-classical approaches have emerged as promising solutions to overcome the limitations of exclusively quantum algorithms. These methods combine the nonlinearity, maturity, and stability of classical computation with the enhanced capacity, expressivity, and inherent parallelism of quantum systems. While classical components can efficiently handle pre-/post-processing, optimization, and error stabilization, quantum layers can capture intricate patterns, solve sub-problems, and accelerate specialized computations.

Within this hybrid paradigm, Quantum Neural Networks (QNNs), also referred to as parameterized quantum circuits (PQC), have demonstrated universal approximation capabilities for continuous functions when implemented with sufficient depth, width, and well-designed quantum feature maps. For instance, Salinas et al.~\cite{PhysRevA.104.012405} showed that single-qubit networks with iterative input reuploading can approximate bounded complex functions, while Goto et al.~\cite{goto2021universal} established universal approximation theorems for quantum feedforward networks. Schuld et al.~\cite{schuld2021effect} revealed that the outputs of parameterized quantum circuits (with suitable data embedding and circuit depth) can be expressed as truncated Fourier series, thereby providing theoretical support for the universal approximation ability of QNNs. Liu et al.~\cite{liu2021rigorous} provided formal evidence of quantum advantage in function approximation. 
Killoran et al~\cite{killoran2019continuous} demonstrate the universality of the continuous variable quantum models by showing that many classical neural networks can be effectively emulated on a photonic quantum computer.
These developments not only support applications in solving PDEs and optimization tasks but also extend to other computational tasks such as quantum chemistry~\cite{cao2019quantum}, material sciences~\cite{bauer2020quantum, nachman2021quantum}, and pattern recognition~\cite{trugenberger2002quantum}, to name a few.

A well-known example of this integrated approach is the Variational Quantum Algorithms (VQA) or Variational Quantum Circuits (VQC). These algorithms employ parameterized quantum circuits with tunable gates optimized through classical feedback loops. 
VQA implementations for solving PDEs include the Variational Quantum Eigensolver (VQE)~\cite{peruzzo2014variational,mcclean2016theory}, Variational Quantum Linear Solver (VQLS)~\cite{bravo2023variational}, and Quantum Approximate Optimization Algorithm (QAOA)~\cite{albino2022solving}, as well as emerging methods like Differentiable Quantum Circuits (DQCs)~\cite{kyriienko2021solving,bultrini2023mixed}, quantum kernel methods~\cite{paine2023quantum}, and Quantum Nonlinear Processing Algorithms (QNPA)~\cite{lubasch2020variational}. 

The aforementioned methods primarily utilize qubit-based quantum computing, which represents the predominant paradigm in quantum computing literature. However, continuous-variable quantum computing (CVQC)~\cite{braunstein2005quantum} offers an alternative computational model that leverages continuous degrees of freedom, such as quadratures and electromagnetic field momentum. For consistency, we refer to qubit-based models as discrete-variable (DV) circuit models throughout this work.

Recently, VQA have been extended to Physics-informed quantum neural networks (PIQNN)~\cite{setty2023physics, dehaghani2024hybrid, trahan2024quantum,sedykh2024hybrid, leong2025hybrid}, broadening their applicability across physics and engineering domains. These hybrid methods offer several distinct advantages when implemented within the PINN framework. 

Firstly, these models enable the flexible configuration of inputs and outputs independent of qubit/qumode counts or quantum layer depth, making them particularly effective for scenarios where the input dimension (e.g., temporal and spatial coordinates) differs from output dimensions (e.g., dependent variables like velocity, pressure, and force).
Secondly, hybrid approaches facilitate increased trainable parameter counts without the need for deeper quantum circuits, additional qubits, or higher cutoff dimensions in CV quantum systems. These factors are particularly crucial in the NISQ era, where hardware constraints impose strict limits on circuit depth and coherence times, making deeper quantum networks computationally expensive and impractical~\cite{preskill2018quantum, bharti2022noisy}. 
Thirdly, these qualities are especially advantageous for PINNs, where shallow network architectures are favored to optimize the balance between computational efficiency and solution accuracy~\cite{krishnapriyan2021characterizing, wang2022approximation, shin2023error, farea2024:arXiv}.
Finally, a substantial benefit of such hybrid models is the integration of automatic differentiation within both classical and quantum networks. This capability, supported by contemporary software packages like PennyLane~\cite{bergholm2018pennylane} and TensorFlow Quantum~\cite{broughton2020tensorflow}, enables the automatic computation of derivatives as variables are directly incorporated into the computational graph. This approach simplifies PINN implementation and facilitates the efficient handling of complex physics-informed loss functions and higher-order derivatives.

Our work is related to the approaches in~\cite{sedykh2024hybrid,dehaghani2024hybrid,trahan2024quantum,leong2025hybrid}, but with several key differences. While HQPINN~\cite{sedykh2024hybrid} focuses specifically on computational fluid dynamics in 3D Y-shaped mixers using a fixed quantum depth with an infused layer structure, the QPINN~\cite{dehaghani2024hybrid} integrates a dynamic quantum circuit combining Gaussian and non-Gaussian gates with classical computing methodologies to solve optimal control problems. In addition, the model proposed in~\cite{trahan2024quantum} investigates both purely quantum and hybrid PINNs, highlighting parameter efficiency but relying on limited PQC networks and addressing primarily simple PDEs. Recently proposed HQPINN~\cite{leong2025hybrid}, inspired by the Fourier neural operator (FNO)~\cite{li2021fourier}, introduces a parallel hybrid architecture to separately model harmonic and non-harmonic features before combining them for the output. \rev{ The present work addresses general PDE solving with multiple quantum circuit configurations and explicit emphasis on parameter efficiency.}

%---------------------------------------------------------------------------------------------------------%
\section{Background}
\label{sec:background}
%---------------------------------------------------------------------------------------------------------%

%---------------------------------------------------------------------------------------------------------%
\subsection{Physics Informed Neural Network (PINNs)}
%---------------------------------------------------------------------------------------------------------%

\rev{In the present work, we consider a general partial differential equation of the form:
\begin{align}
    \label{eq:general_pde_eq}
    \mathcal{D}[u(\mathbf{x});\alpha] &= f(\mathbf{x}) , \quad \mathbf{x} \in \Omega, 
    \\
    \mathcal{B}_k[u(\mathbf{x})] &= g_k(\mathbf{x}) , \quad \mathbf{x} \in \Gamma_k \subset \partial \Omega ,\quad k =1,2,\hdots ,n_b \notag
\end{align}
\noindent where, $\mathcal{D}$, $\mathcal{B}_k$ denote arbitrary differential and boundary/initial operators, respectively. The term $f(\mathbf{x})$ refers to the source term or forcing function in the PDE. The functions $g_k(\mathbf{x})$ refer to a set of boundary/initial functions, $\alpha$ is a set of parameter vector or coefficient and $u(\mathbf{x})$ is the solution of the differential equation given input $\mathbf{x}$  in the solution domain $\Omega$ with boundary $\partial \Omega$.
}

PINNs~\cite{raissi2019physics} provide a framework to embed governing physical laws directly into neural network architectures.
The PINN loss function combines terms representing governing equations and initial/boundary conditions designed to minimize deviations from physical laws. More formally, the PINN loss function can be defined as: 

\begin{equation} 
    \begin{split}
    \mathcal{L}(\theta) =& \arg\min_{\theta} \sum_{k=1}^{n} \lambda_k \mathcal{L}_k(\theta) \\
    =& \arg\min_{\theta} \left( \lambda_1 \mathcal{L}_1(D[u_{\theta}(\mathbf{x}); \mathbf{\alpha}] - f(\mathbf{x}))
    +\sum_{k=2}^{n_b} \lambda_k \mathcal{L}_k(B[u_{\theta}(\mathbf{x})] - g_k(\mathbf{x})) \right),
    \end{split}
 \end{equation}

\noindent where, $n_b$ denotes the total number of loss terms, \rev{$\mathcal{L}_k$ are the individual loss functions for each constraint with weighting coefficients $\lambda_k$}, and $\theta$ represents the trainable parameters.  

While the initial and boundary conditions guarantee a unique solution by satisfies predefined constraints at specific spatial or temporal locations,  the physics loss acts as a regularizer, guiding the network to learn physically consistent solutions and preventing overfitting to sparse or noisy data. The balance between these loss components is crucial for achieving accurate solutions, as improper weighting can lead to underfitting the initial/boundary conditions or failing to capture the system dynamics correctly. In the present work adopts soft imposition of the boundary and initial conditions using fixed weights with the best empirical results. \rev{However, various weighting methods have been proposed in the literature using different criteria, such as 
gradient information~\cite{wang2021understanding,vemuri2023gradient,Liu2024ConFIG}, 
loss-term history statistics~\cite{Heydari2020,bischof2021multi}, 
training dataset information~\cite{mcclenny2020self,li2022dynamic,sahli2020physics}, or hard imposition of initial and boundary conditions~\cite{ren2022phycrnet,norambuena2024physics},to name a few.}

%---------------------------------------------------------------------------------------------------------%
\subsection{Continuous Variable (CV) QNN}
\label{sec:cv-qnn}
%---------------------------------------------------------------------------------------------------------%

In photonic quantum computing, CV computation addresses quantum states with continuous degrees of freedom, such as the position ($x$) and momentum ($p$) quadratures of the electromagnetic field. 
\rev{Within this framework, qumodes represent the fundamental units of the CV quantum systems, analogous to qubits in discrete-variable systems. Unlike qubits, qumodes encode information in the continuous degrees of freedom of electromagnetic field modes, rather than discrete two-level states.}
The computational space is, in principle, unbounded, involving infinite-dimensional Hilbert spaces with continuous spectra. To make these simulations computationally feasible, a cutoff dimension is employed to limit the number of basis states used for approximation. 
The computational state space is defined as $r = m^n$, with $n$ qumodes and cutoff dimension $m$. For example, with three qumodes and a cutoff dimension of 2, the state space is 8-dimensional. 
Ideally, higher cutoff dimensions (up to a certain threshold) yield more precise results but at the cost of increased computational complexity.

In this work, we adhere to the affine transformation introduced in \cite{killoran2019continuous}. The proposed circuit ansatz consists of a sequence of operations that manipulate quantum states in the phase space.  Each layer of the CV-Circuit QNN includes the following key components:

\begin{enumerate}
    \item \textbf{Interferometers (\(U_1(\psi_1, \phi_1)\) and \(U_2(\psi_2, \phi_2)\))}: Linear optical interferometers perform orthogonal transformations, represented by symplectic matrices. These transformations adjust the state by applying rotation and mixing between modes, corresponding to beam splitting and phase-shifting operations. 
    \item \textbf{Squeezing Gates (\(S(r, \phi_s)\))}: These gates scale the position and momentum quadratures independently, introducing anisotropic adjustments to the phase space. Encoding sharp gradients using squeezing gates is particularly appealing for PDE solvers, as it allows fine-tuning of local characteristics \cite{sedykh2024hybrid}.
    \item \textbf{Displacement Gates (\(D(\alpha, \phi_d)\))}: These gates shift the state in phase space by adding a displacement vector \(\alpha\), translating the state globally without altering its internal symplectic structure.  
    \item \textbf{Non-Gaussian Gate (\(\Phi(\lambda)\))}: To introduce nonlinearity, a non-Gaussian operation, such as a cubic phase, Kerr, and cross-Kerr gates, is applied. This step enables the neural network to approximate nonlinear functions and enhances the ability to solve complex, nontrivial PDEs.
\end{enumerate}

\noindent  
These components together mimic the classical affine transformation such that:
\begin{equation}
    L(\ket{x}) = \ket{\Phi(Mx + b)} =  \Phi(D \circ U_2 \circ S \circ U_1),
\label{eq:affine_transform}
\end{equation}

\noindent where, \(M\) represents the combined symplectic matrix of the interferometers and squeezing gates, \(x\) is the input vector in the phase space, $b$ is the displacement vector with nonlinear transformation \(\Phi\). 

In addition to the building blocks above, CV-Circuit QNNs offer flexibility in measurement schemes and parameterization strategies. Measurement can be performed in either the number basis, which projects the state onto discrete photon-number states, or the quadrature basis, which yields continuous-valued outcomes corresponding to position or momentum observables. The parameterization of unitary operations may either be full (including both magnitude and phase) or restricted to magnitude-only (by setting phase parameters \(\phi\) to zero), which can improve trainability and mitigate optimization challenges at the cost of reduced expressivity.

Although such a model exhibits intriguing theoretical properties, their practical implementation in real-world problem solving seems to suffer from fundamental stability issues. This is mainly due to their inherent sensitivity and reliance on continuous degrees of freedom, which complicate optimization compared to DV-based approaches (discussed in detail in section~\ref{sec:dv-qnn}). This sensitivity stems from several factors, including the precision required for state preparation, noise from decoherence, and the accumulation of errors in variational quantum circuits. Furthermore, the transformations involved in CV quantum circuits are prone to numerical instabilities, particularly when optimized using gradient-based methods. Although foundational studies about the barren plateau problem, such as~\cite{mcclean2018barren,cerezo2021cost,holmes2022connecting}, primarily address DV circuit systems, the analysis indicates that similar issues can arise in CV quantum systems. In particular, the infinite-dimensional Hilbert space and the continuous nature of parameterization can exacerbate the barren-plateau effects.

%---------------------------------------------------------------------------------------------------------%
\subsection{Discrete Variable (DV) QNN}
\label{sec:dv-qnn}
%---------------------------------------------------------------------------------------------------------%

DV-Circuit quantum models do not inherently provide affine transformations as a core feature like CV-based QNNs discussed in~\ref{sec:cv-qnn}. However, they can achieve affine-like behaviour through parameterized single-qubit rotations, controlled operations, and phase shifts for amplitude adjustments. Therefore, the affine transform of equation (\ref{eq:affine_transform}) can be reinterpreted to align with their architecture such that:

\begin{enumerate}
    \item $\mathbf{ Mx + b }$: can be achieved through a combination of parameterized quantum gates:\\
    - \( U_1, U_2 \): A unitary operator that can be represented by parameterized single-qubit rotations, such as \( RX(\theta), RY(\theta) \), or \( RZ(\theta) \).\\
    - \( S \): A sequence of controlled gates (e.g., CNOT or CZ) responsible for introducing entanglement, which enables the representation of correlated variables and interactions.\\
    - \( D \): Data encoding circuits that map classical data into quantum states through techniques such as basis embedding, amplitude embedding, or angle embedding.

    \item \textbf{Non-linear Activation (\(\Phi\))}:
    DV-Circuit QNNs lack intrinsic quantum nonlinearity due to the linearity of quantum mechanics. However, classical measurements or a combination of classical measurements and non-linear activation functions (such as Tanh) can be applied to the output between quantum layers~\cite{tacchino2020quantum,mangini2021quantum}.
    
\end{enumerate}

%---------------------------------------------------------------------------------------------------------%
\section{Quantum-Classical Physics-Informed Neural Networks (QCPINN)} 
\label{sec:QCPINN}
%---------------------------------------------------------------------------------------------------------%

\begin{figure}[t]
\centering
\ifIOP
    \includegraphics[width=1\textwidth,trim=0.0cm 0.0cm 0.0cm 0.0cm, clip]{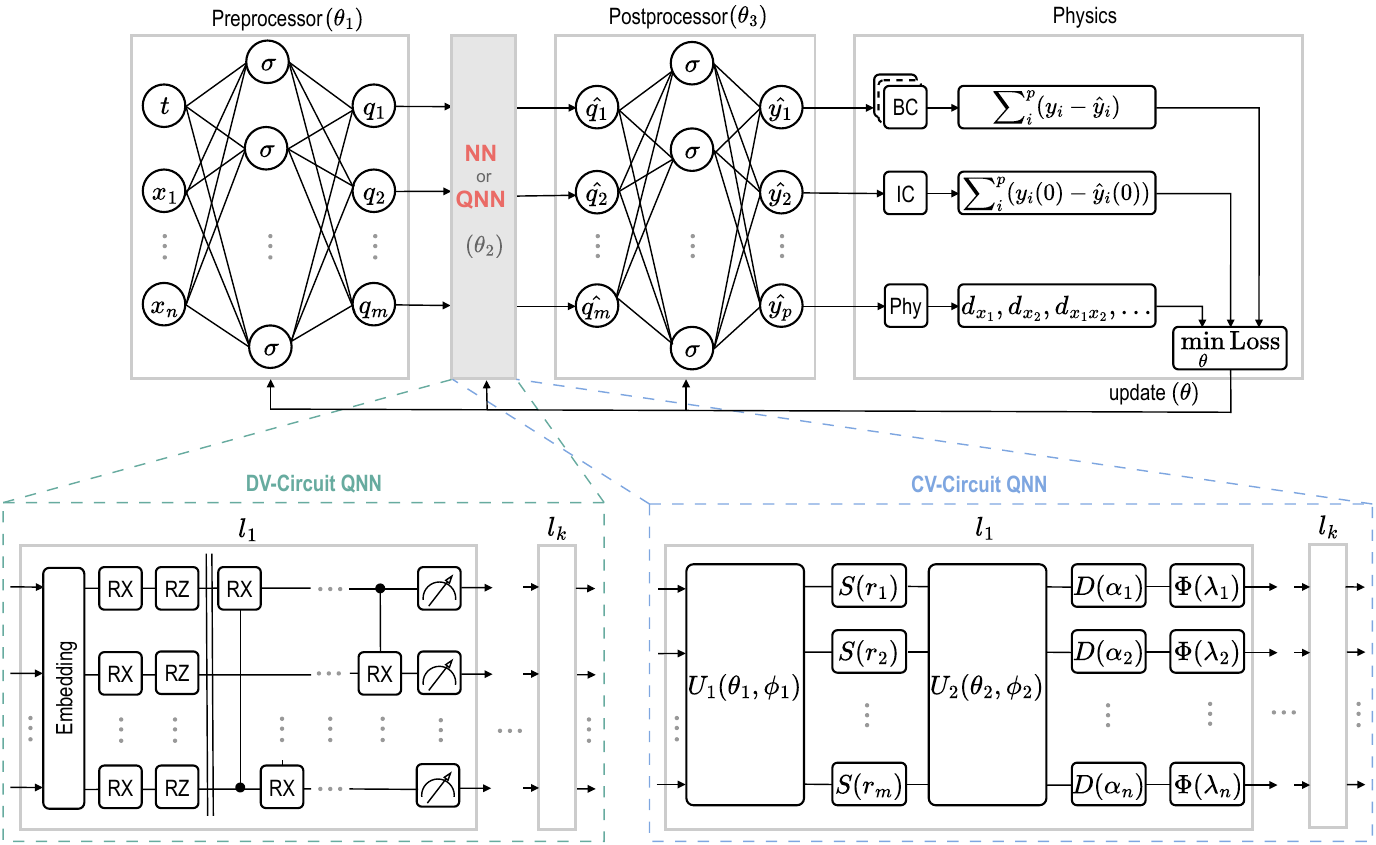} 
\fi
\ifARXIV
    \includegraphics[width=0.8\textwidth,trim=0.0cm 0.0cm 0.0cm 0.0cm, clip]{figures/qcpinn.pdf} 
\fi
       
\caption{
    Overview of the proposed QCPINN architecture, illustrating its hybrid structure consisting of three main components: (a) a classical preprocessor for input encoding, (b) the core processing unit using either a quantum neural network (QNN) or classical neural network (NN), and (c) a classical postprocessor for decoding the QNN or NN outputs. The QNN is implemented as either a CV or DV circuit with $k$ layers. The architecture integrates a loss function to concurrently minimize PDE residuals and enforce compliance with initial and boundary conditions, thereby ensuring adherence to physical constraints during training. The various configurations of the CV and DV circuits are detailed in Table~\ref{tab:qcpinn_config}. 
}
    
\label{fig:architecture}
\end{figure}

\begin{table}[t]
\centering
\caption{
Various configurations of the QCPINN, along with the list of PDEs used for testing. \rev{Each configuration was evaluated against the Helmholtz and lid-driven Cavity cases, with the top-performing setups subsequently selected for further assessment on additional scenarios, including Wave, Klein-Gordon, and Convection-diffusion problems.}
}
\label{tab:qcpinn_config}
\scriptsize

\begin{tabular}{l l p{6cm} | l} 
    \toprule

    \multirow{2}{*}{\textbf{QNN}}& \multicolumn{2}{c|}{\textbf{Configurations}}& \multirow{2}{*}{\textbf{PDEs}}\\
    \cmidrule(lr){2-3}

    & \textbf{Type} & \textbf{Options} & \\
    
    \midrule
    
    \multirow{3}{*}{CV-Circuit}& Measurement schemes& Number, Quadrature& \multirow{5}{*}{\makecell[l]{ \textit{(tested on all configurations)} \\ Helmholtz, Lid-driven Cavity\\ \\ \textit{(tested on selected configurations)} \\ Wave, Klein-Gordon, \\Convection-diffusion}} \\ 

    \cmidrule(){3-3}

    & Nonlinear operations & Kerr, Cubic phase, Cross-kerr & \\

    \cmidrule(){3-3}
    
    & Parameterization & \makecell[l]{Phase-free (magnitude), Full (magnitude \\ and phase) (see Algorithm~\ref{alg:cv-qcpinn})}& \\ 
    
    \cmidrule(){1-3}
    
    \multirow{2}{*}{DV-Circuit} 

    & Embeddings & Amplitude, Angle & \\ 
    
    \cmidrule(){3-3}

    & Topology & \makecell[l]{Alternate, Cascade, Cross-mesh, Layered \\ (see Fig.~\ref{fig:dv_circuits} and Table~\ref{tab:topology_comparison})} & \\ 
    
    \bottomrule
    
    \end{tabular}
\end{table}

\begin{figure}[t]
\centering
\ifIOP
\begin{tabular}{cc}
    \includegraphics[width=0.40\columnwidth, trim={0cm 0cm 0cm 0cm}, clip]{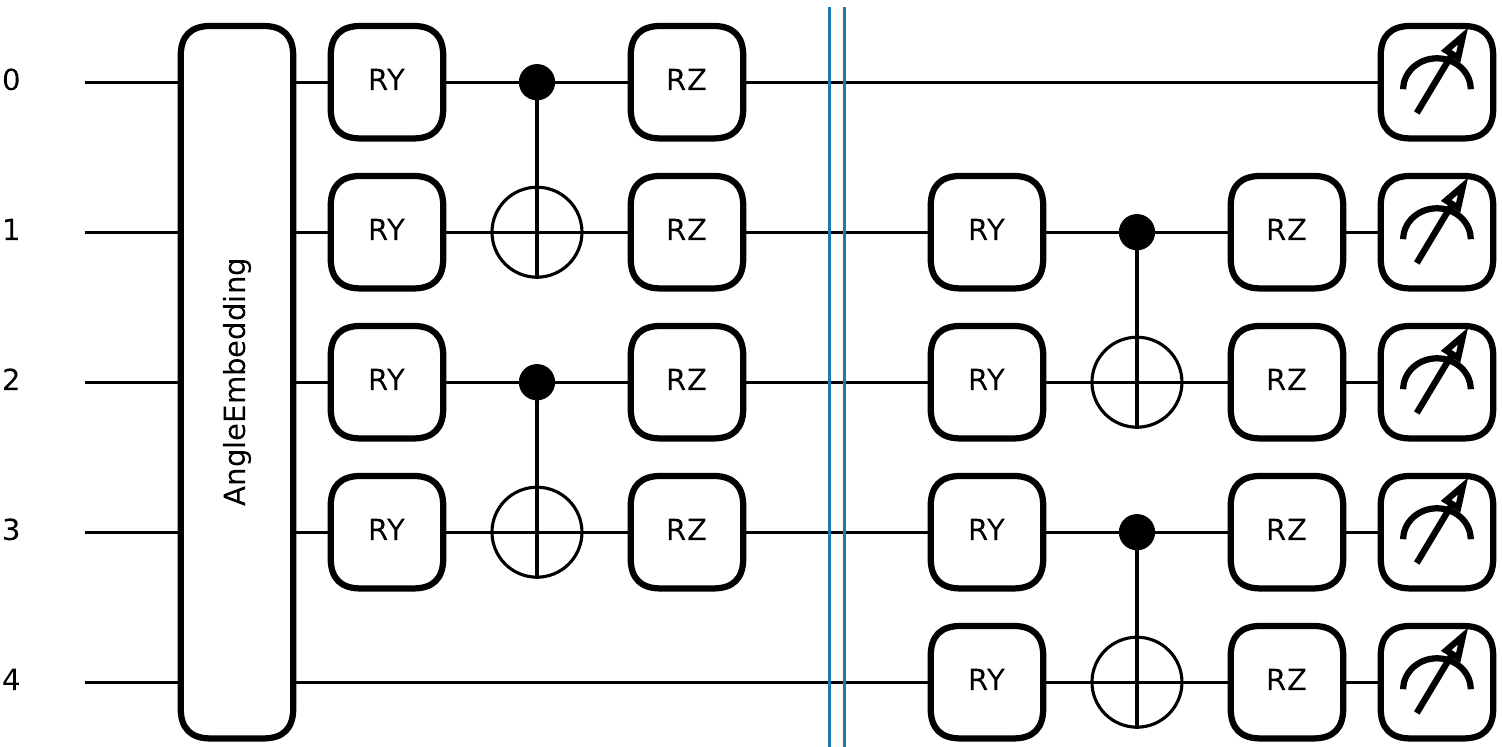} &
    \includegraphics[width=0.45\columnwidth, trim={0cm 0cm 0cm 0cm}, clip]{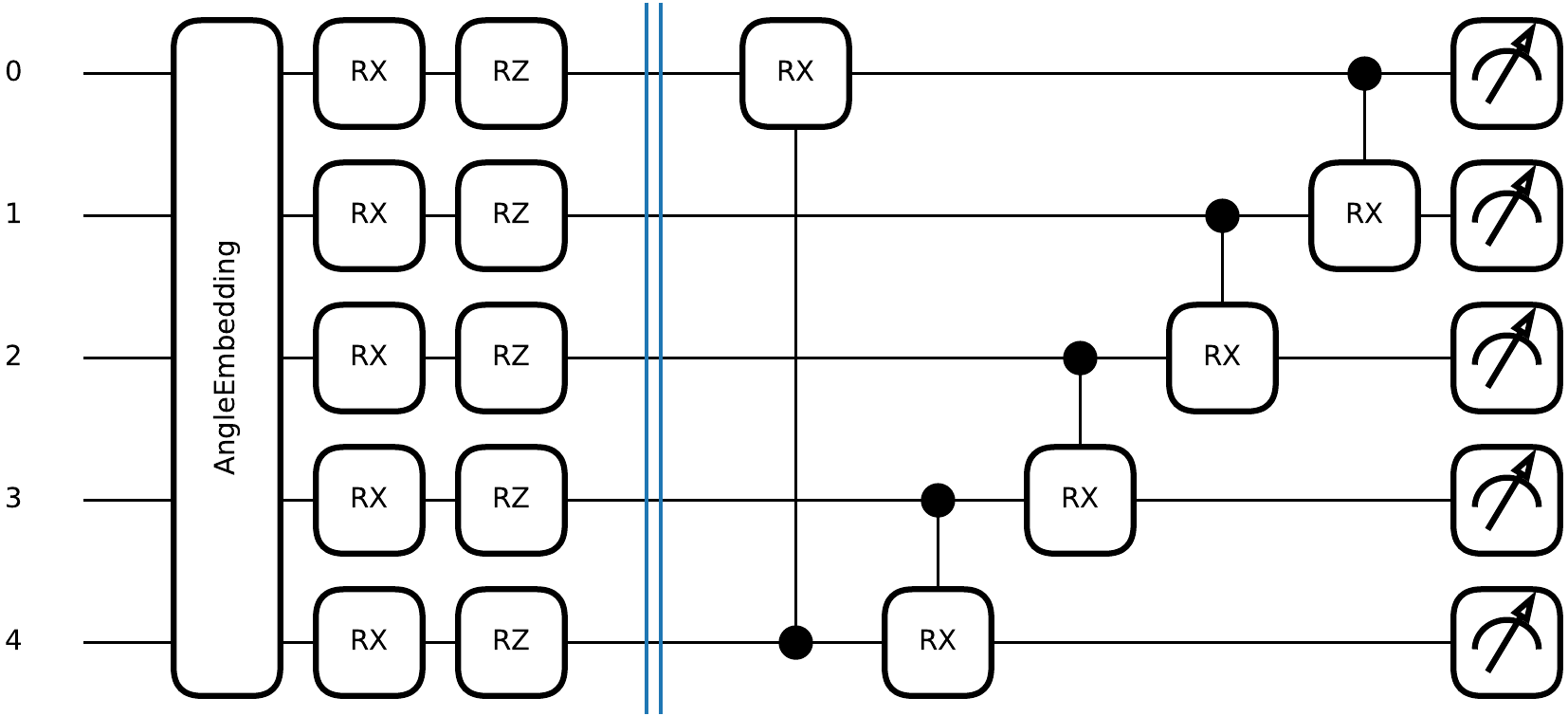} \\
    \footnotesize{(a) Alternate} & \footnotesize{(b) Cascade} \\
    \multicolumn{2}{c}{\includegraphics[width=1.0\columnwidth, trim={0cm 0cm 0cm 0cm}, clip]{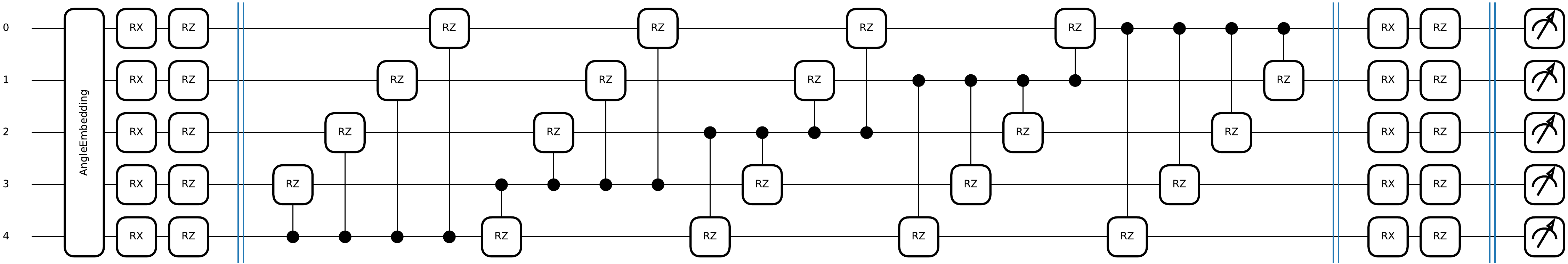}} \\
    \multicolumn{2}{c}{\footnotesize{(c) Cross-mesh}} \\
    \multicolumn{2}{c}{\includegraphics[width=0.55\columnwidth, trim={0cm 0cm 0cm 0cm}, clip]{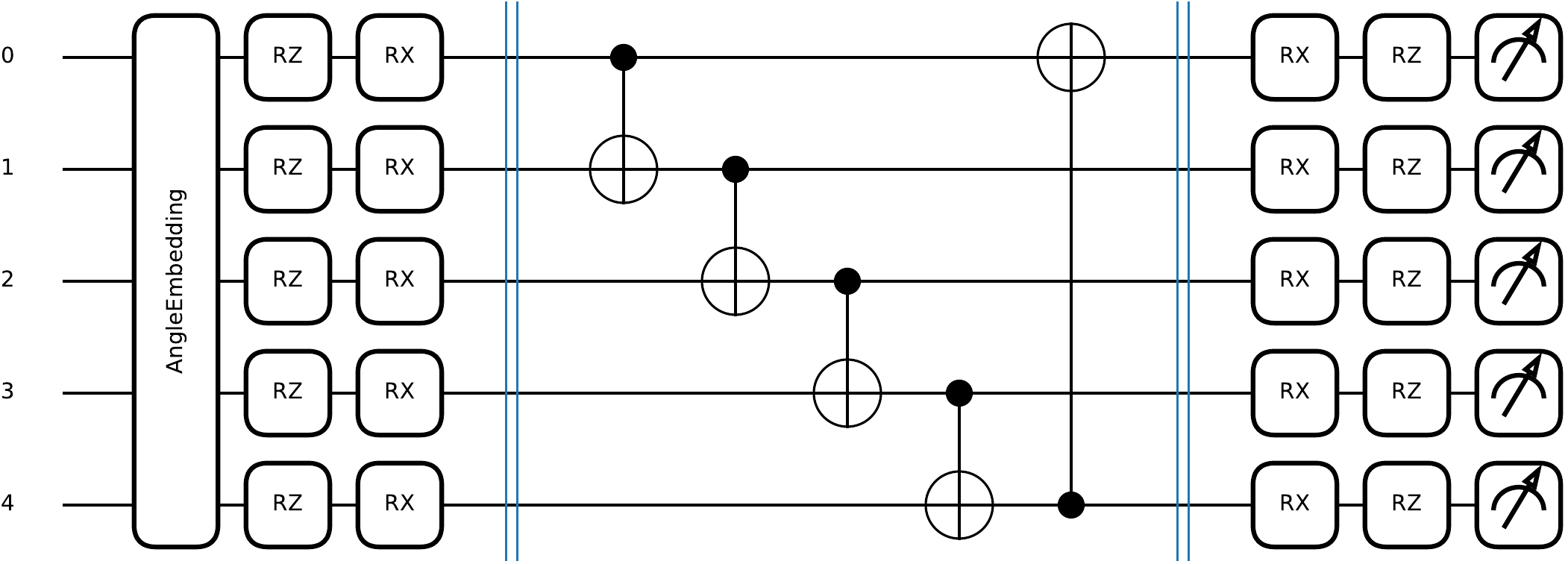}} \\
    \multicolumn{2}{c}{\footnotesize{(d) Layered}}
\end{tabular}
\fi

\ifARXIV
\begin{tabular}{cc}
    \includegraphics[width=0.35\columnwidth, trim={0cm 0cm 0cm 0cm}, clip]{figures/alternate.pdf} &
    \includegraphics[width=0.40\columnwidth, trim={0cm 0cm 0cm 0cm}, clip]{figures/cascade.pdf} \\
    \footnotesize{(a) Alternate} & \footnotesize{(b) Cascade} \\
    \multicolumn{2}{c}{\includegraphics[width=0.90\columnwidth, trim={0cm 0cm 0cm 0cm}, clip]{figures/cross_mesh.pdf}} \\
    \multicolumn{2}{c}{\footnotesize{(c) Cross-mesh}} \\
    \multicolumn{2}{c}{\includegraphics[width=0.60\columnwidth, trim={0cm 0cm 0cm 0cm}, clip]{figures/layered.pdf}} \\
    \multicolumn{2}{c}{\footnotesize{(d) Layered}}
\end{tabular}
\fi

\caption{
    DV-Circuit topologies evaluated in this study, shown with five qubits and angle embedding as an example. Square boxes denote parameterized single-qubit rotation gates, and lines with dots indicate entangling operations (e.g., CNOT gates). The characteristics of these circuits are shown in Table~\ref{tab:topology_comparison}. 
}
\label{fig:dv_circuits}
\end{figure}

Fig.~\ref{fig:architecture} illustrates a high-level overview of our proposed QCPINN architecture. The preprocessor consists of a two-layer classical neural network. The first layer performs a linear transformation to modify the input dimension, followed by a nonlinear activation function to introduce classical non-linearity. The second layer subsequently maps the transformed inputs to the QNN layer (or classical NN layer used for comparison). 
The postprocessor layer mirrors the structure of the preprocessor, maintaining consistent dimensions and ensuring a symmetric transformation between classical and quantum representations. This design provides flexibility in encoding classical inputs to quantum inputs and decoding quantum outputs back to classical outputs. Additionally, it manages input-output dimensions and regulates quantum feature representations, all while ensuring the higher-order differentiability necessary for residual loss in the proposed QCPINN model.

%---------------------------------------------------------------------------------------------------------%
\subsection{CV-Circuit QCPINN}
%---------------------------------------------------------------------------------------------------------%
\label{sec:CV-Circuit-QCPINN}
The CV-Circuit QCPINN consists of three primary components: a classical preprocessor network, a CV-Circuit quantum network (the QNN layer of Fig.~\ref{fig:architecture}), and a classical postprocessor network. The quantum network is adapted from~\cite{killoran2019continuous} and is described in Algorithm~\ref{alg:cv-qcpinn}. Each layer of the quantum network executes a sequence of quantum operations. 

Table~\ref {tab:qcpinn_config} presents various configurations of our CV-Circuit studied to identify a suitable parameterization scheme, nonlinear operation, and measurement approach. 
For measurement schemes, we evaluated both number and quadrature.
For parameterization, we examined phase-free using only magnitude parameters and full parameterization using both magnitude and phase parameters. 
For nonlinear operations, we explored Kerr nonlinearity, cubic phase gates, and cross-Kerr interactions. 
Additionally, the quantum state is measured, typically in either the quadrature basis $\langle \hat{q}_i \rangle$ or the Fock number basis $\langle \hat{n}_i \rangle$, to yield classical outputs for the postprocessing layers. Please refer to section~\ref{sec:cv-qnn} for additional details.

Structuring the network in this way results in a circuit depth of $(2n+3)L$, while the number of trainable quantum parameters increases to $\mathcal{O}(n^2 L)$, with $n$ qumodes and $L$ number of layers.

%---------------------------------------------------------------------------------------------------------%
\subsection{DV-Circuit QCPINN}
\label{sec:dv-circuit-qcpinn}
%---------------------------------------------------------------------------------------------------------%

\begin{table}[]
    \centering
    \caption{
    Characteristics of the topologies used in DV-Circuits in Fig.~\ref{fig:dv_circuits}, where $n$ is the number of qubits, and $L$ is the number of layers. 
    }
    \label{tab:topology_comparison}
    \scriptsize
    \begin{tabular}{l c c c c}
        \toprule
        \textbf{Topology} & 
        \textbf{\makecell{Circuit\\Depth}} & 
        \textbf{\makecell{Circuit\\Connectivity}} & 
        \textbf{\makecell{Number of\\Parameters}} & 
        \textbf{\makecell{Number of\\two-qubit gates}}  \\
        \midrule
        Alternate &  $6L$& Nearest-neighbor & $4(n-1)L$ & $(n-1)L$  \\
        Cascade & $(n + 2)L$ & Ring topology & $3nL$ & $nL$  \\
        Cross-mesh & $(n^2 - n + 4)L$ & All-to-all & $(n^2 +4n)L$  & $(n^2 -n)L$  \\
        Layered &  $6L$& Nearest-neighbor & $4nL$ & $ (n-1)L$  \\
        \bottomrule
    \end{tabular}
\end{table}

Similar to the QCPINN (CV-Circuit) discussed in~\ref{sec:CV-Circuit-QCPINN}, the DV-Circuit QCPINN  consists of three primary components: a classical preprocessor network, a quantum network (the QNN layer of Fig.~\ref{fig:architecture}), and a classical postprocessor network. 
The quantum network is a custom DV-Circuit, which executes quantum operations on the preprocessed data. 
Table~\ref {tab:qcpinn_config} presents various configurations of our DV-Circuit studied to identify a suitable embedding scheme and topology. 
Fig.~\ref{fig:dv_circuits} illustrates four proposed ansätze named after their topology: Alternate, Cascade, Cross-mesh, and Layered. These circuits are adapted from literature, including~\cite{farhi2018classification, sim2019expressibility, jaderberg2020minimum, setty2023physics}. \rev{All square boxes in the figure denote parameterized single-qubit or controlled-rotation gates (RX, RY, RZ, CRX, CRZ), which are optimized during training.}

Table~\ref{tab:topology_comparison} presents features of the DV-Circuits across the key descriptors, circuit depth, connectivity, number of parameters, and the number of two-qubit gates. 
\rev{Here, circuit depth refers to the longest computational path in the circuit from the data input to the output, which represents the maximum number of logically dependent gates executed on a single qubit. 
This metric is critical as deeper circuits require more computational resources and may incur numerical errors that can degrade gradient quality during neural network optimization.
}
The chosen circuits agree with the affine transformation discussed in section~\ref{sec:background} and adhere to the hardware-efficient ansatz (HEA)~\cite{mouton2024deep} defined as:
\begin{align}
\label{eq:HEA}
U(\psi) \;=\; \prod_{k} U_k(\psi_k) \, W_k,
\end{align}
where, each layer $k$ alternates between single-qubit parameterized rotations, $U_k(\psi_k)$, and an entangling operation, $W_k$. 

The \textbf{Cross-mesh} ansatz implements an all-to-all connectivity scheme, facilitating extensive interaction between qubits, which enhances its ability to explore the Hilbert space. The circuit depth scales as $\mathcal{O}(n^2 - n + 4)L$, while the number of trainable parameters is $(n^2 + 4n)L$. The use of global entanglement fosters strong correlations between qubits; however, it results in larger circuit depth, resource demands, and training difficulties~\cite{holmes2022connecting}. This circuit's total number of two-qubit gates is also significantly higher than in other ansätze. 

The \textbf{Cascade} ansatz balances expressivity and hardware efficiency by employing a ring topology for qubit connectivity. This design facilitates efficient entanglement while keeping the circuit depth manageable at $(n+2)L$. The circuit contains $3nL$ trainable parameters, making it more efficient than Cross-mesh while maintaining reasonable expressivity. Incorporating controlled rotation gates ($CRX$) improves entanglement compared to CNOT gates and increases the circuit's capacity to explore the Hilbert space. Furthermore, the requirement for two-qubit gates is fewer than that of Cross-mesh, thus lowering the overall computational cost.

The \textbf{Layered} ansatz features a structured design that includes alternating rotation and entangling layers. Each qubit undergoes parameterized single-qubit rotations, typically utilizing $RX$ and $RZ$ gates. Entanglement is achieved through nearest-neighbor CNOT gates. Compared to the Cross-mesh's all-to-all connectivity or the Cascade's ring topology, the Layered architecture represents a compromise between expressivity and hardware efficiency. 

\rev{
Among the evaluated DV topologies, the Alternate ansatz employs alternating layers of parameterized single-qubit rotations followed by nearest-neighbor CNOT gates, ensuring effective entanglement while maintaining a manageable circuit depth of $6L$ and  $4(n-1)L$ trainable parameters. Although we did not compute entanglement entropy explicitly, this layout is known to generate strong bipartite and multipartite entanglement in shallow hardware-efficient circuits \cite{sim2019expressibility, cerezo2021variational, holmes2022connecting}.
}

Compared to the QNN models discussed above, we can see that the classical feed-forward neural networks (NNs) exhibit a more flexible scaling pattern, with trainable parameters scale as $\mathcal{O}(h_ih_{i+1})$, where $h_i$ and $h_{i+1}$ denote the number of neurons in two consecutive layers $l_i$ and $l_{i+1}$  with independent selection of the neurons. As a result, while classical models depend on deep architectures and larger parameter counts, quantum networks achieve expressivity by utilizing intrinsic quantum correlations and exponential scaling of the Hilbert space. \rev{Consequently, incorporating a QNN between two consecutive classical layers, like the structure shown Fig.~\ref{fig:architecture}, with number of qubits much less than the number of neurons can reduce the number of parameters from, say, $\mathcal{O}(h_{i}h_{i+1})$ to $\mathcal{O}(h_{i}n_{i} + q_{\text{params}} + n_{i}h_{i+1})$ where $n_{i}$ is the number of qubits and $q_{\text{params}}$ represents the quantum circuit parameters.
}
%---------------------------------------------------------------------------------------------------------%
\section{Implementation and Training}
\label{sec:implementation}
%---------------------------------------------------------------------------------------------------------%

%---------------------------------------------------------------------------------------------------------%
\subsection{CV-Circuit QCPINN}
\label{sec:cv-training}
%---------------------------------------------------------------------------------------------------------%

We implemented the preprocessor and the postprocessor (as in Fig.~\ref{fig:architecture}) with one hidden layer with 50 neurons and Tanh activation function. 
We used the PennyLane framework~\cite{bergholm2018pennylane} with the Strawberry Fields Fock backend~\cite{killoran2019strawberry} for the CV-Circuit QCPINN implementation. Each CV-Circuit layer was configured for two qumodes with a cutoff dimension of 20. Quantum parameters were initialized using controlled random distributions, with small values for active parameters (e.g., $r$, $d$) to ensure numerical stability. 
Classical neural network layers were initialized using Xavier initialization\rev{~\cite{glorot2010understanding}}. During training, either position quadrature (X) or number operator expectation measurements are performed on each qumode to extract continuous-valued features from the quantum state. 

\rev{As CV-circuits rely on continuous degrees of freedom, their operations are inherently more sensitive to parameter scaling than those in DV systems. This sensitivity often manifests as instability during training, particularly when parameters grow large or when nonlinear transformations amplify fluctuations in the quantum state}. For instance, squeezing operations are highly sensitive to large values, as excessive squeezing amplifies quadrature fluctuations, which increase photon number variance and truncation errors in Fock-space representations~\cite{lvovsky2015squeezed}. Similarly, displacement operations are constrained to prevent quantum states from exceeding computational cutoffs, as this can introduce simulation errors~\cite{ferraro2005gaussian}. Likewise, the nonlinear operations such as the Kerr gate can cause phase shifts that scale quadratically with photon numbers, making them highly sensitive to large values~\cite{drummond1980quantum}. 

Therefore, we used a learning rate of $1.0 \times 10^{-4}$ to address the model’s sensitivity to gradient updates. A ReduceLROnPlateau scheduler was implemented with a patience of 20 epochs, decreasing the learning rate by a factor of 0.5 when no improvement was observed, down to a minimum of $1.0 \times 10^{-6}$. Additionally, we applied gradient clipping to prevent exploding gradients, restricting their norm to a maximum of 0.1. We also scaled the training parameters by small factors to further stabilize the CV transformations.
While these settings can alleviate the instability and prevent divergence in quantum state evolution, they can also introduce the vanishing gradient problem~\cite{glorot2010understanding}.

%---------------------------------------------------------------------------------------------------------%
\subsection{DV-Circuit QCPINN}
%---------------------------------------------------------------------------------------------------------%

Similar to the CV-Circuit QCPINN, the preprocessor and the postprocessor consist of one hidden layer with 50 neurons and the Tanh activation function.
We used Adam optimizer with a learning rate of 0.005 to train the DV-Circuit QCPINN. A learning rate scheduler, ReduceLROnPlateau, is employed to adjust the learning rate when the loss plateaus dynamically. Specifically, the scheduler reduces the learning rate by a factor of 0.9 if no improvement is observed for 1000 consecutive epochs, promoting stability and convergence. Furthermore, the training function applies gradient clipping to prevent exploding gradients, clipping them according to their norm to ensure they do not exceed a unity value. During training, Pauli-Z expectation measurements are performed on each qubit to extract features from the quantum state.

\rev{In all DV-Circuit QCPINN  experiments, we set the number of shots to None to enable analytic gradient computation through backpropagation with PennyLane. This configuration activates the deterministic simulation mode, allowing continuous gradients to flow through the quantum circuit and enabling differentiation with backpropagation. Using a finite number of shots would introduce stochastic sampling noise, which would disrupt compatibility with backpropagation and require alternative differentiation methods, such as the parameter-shift rule, that are significantly more computationally expensive.}

%---------------------------------------------------------------------------------------------------------%
\subsection{Classical PINN Models}
\label{sec:classical-pinn-model}
%---------------------------------------------------------------------------------------------------------%

The baseline PINN models consist of three main components: a preprocessing network, a neural network NN, and a postprocessing network (as shown in Fig.~\ref{fig:architecture}). The preprocessor and the postprocessor contain one hidden layer with 50 neurons and the Tanh activation function. The NN has two different configurations: 

\begin{enumerate}
    \label{list:pinn-models}
    \item 
    Model-1 includes two fully connected layers, each containing 50 neurons with a Tanh activation function between layers, and 
    \item 
    Model-2 retains only the preprocessing and postprocessing networks, entirely omitting the NN to simplify the model's structure.
\end{enumerate}

These two models represent different levels of complexity, with model-1 having more layers and, consequently, a higher number of trainable parameters than model-2. We use these baseline PINN models to evaluate and compare the performance of the proposed hybrid models in Fig~\ref{fig:architecture}.

%---------------------------------------------------------------------------------------------------------%
\subsection{PDEs}
%---------------------------------------------------------------------------------------------------------%

We implemented five distinct PDEs, which include the Helmholtz, the time-dependent 2D lid-driven Cavity, the 1D Wave, the Klein-Gordon, and a Convection-diffusion problem. The mathematical formulation, boundary conditions, and corresponding loss function designs for each PDE are detailed in~\ref{app:pde_difinition}.

%---------------------------------------------------------------------------------------------------------%
\subsection{Experiment Setup}
%---------------------------------------------------------------------------------------------------------%
We used PyTorch's automatic differentiation to compute the gradients for both classical and quantum parameters, and the Adam optimizer. The training was carried out with a batch size of 64 and a maximum of 20,000 epochs.
For the Cavity problem, the entire training dataset was generated using the Sobol sequence; in other cases, random sampling was sufficient throughout the training process.
Training settings, including the optimizer and batch size, were kept consistent in QCPINN models to allow for a fair comparison with the baseline PINN. While altering configurations, such as increasing iterations or adjusting the batch size, might improve performance, the aim is to evaluate the models under identical conditions.
Given the near-infinite hyperparameter space, this controlled evaluation ensures a fair assessment of the relative strengths and limitations of each approach. 

In our initial implementation, we observed that the experiments on the CPU were faster than those on the GPU. This is likely due to our study's relatively small batch size, circuit complexity, and limited number of qubits/qumodes: five qubits for DV-Circuit models and two qumodes for CV-based models. Moreover, the quantum backends employed in our experiments are not fully optimized for GPU efficiency. Considering these limitations, we have conducted our experiments on a CPU system with 6148 CPUs (each operating at 2.40 GHz) and 192 GB of RAM.

%---------------------------------------------------------------------------------------------------------%
\rev{
\section{Results}
\label{sec:results}
%---------------------------------------------------------------------------------------------------------%
%
We first investigated the performance of the CV-Circuit QCPINN models across twelve distinct configurations, encompassing two measurement schemes, three nonlinear operations, and two parameterization strategies (as summarized in Table~\ref{tab:qcpinn_config}). These configurations were evaluated on the Helmholtz and lid-driven Cavity problems to assess their stability and convergence characteristics. However, the training behavior of the CV-Circuit models was found to be highly unstable, regardless of the configuration tested. Results are discussed in ~\ref{app:cv-circuit-qcpinn} in detail. In particular, Table~\ref{tab:results_cv} and Fig.~\ref{fig:cv_loss_plots} report the corresponding solution accuracy and training convergence for two representative configurations. 
}

\rev{ 
Following the CV-Circuit analysis, we systematically evaluated the DV-Circuit QCPINN models across a set of configurations comprising two embedding schemes and four circuit topologies. These quantum architectures were benchmarked against baseline PINN models discussed in~\ref{sec:classical-pinn-model} to assess their relative performance in terms of accuracy, convergence, parameter efficiency, time, and memory. We report the results of the best-performing configurations, while the complete set of results for all DV-Circuit configurations is provided in Appendix~\ref{app:dv_circuit_all} for reference.
}

\begin{table}[t]
\centering
\caption{
\rev{
Comparing the results of the best configuration of DV-Circuit QCPINN with PINN.
Reported metrics include $N_p$: the number of trainable parameters; $L_2$: relative error (in \%), and the mean~$\pm$~std across 10 independent runs; $L_\text{final}$: final loss; T: time per iteration (in sec), M: peak memory consumption (in MB). For the Cavity problem, $u$ and $v$: velocity and $p$: pressure, and for other cases, $u$: reference solution and $f$: source term.
For results with additional configurations, see Table~\ref{tab:results_helmholtz}, ~\ref{tab:results_cavity} and ~\ref{tab:additional_cases} in the appendix.
}
}
\label{tab:best-results}

\scriptsize

\begin{tabular}{@{}l
        c@{\hspace{12.0pt}}
        c@{\hspace{4.0pt}}@{\hspace{4.0pt}}
        c@{\hspace{8.0pt}}
        c@{\hspace{8.0pt}}
        c@{\hspace{12.0pt}}
        c@{\hspace{8.0pt}}
        c@{\hspace{8.0pt}}
        c@{\hspace{8.0pt}}
        @{}}
    
    \toprule
        
    \textbf{PDEs}& \multicolumn{2}{c}{\textbf{Model}}& $N_p$& &$L_2$ &$\mathcal{L}_{\text{final}}$ & T& M\\
    
    \midrule

     \multirow{4}{*}{\makecell[l]{Helmholtz\\Eq.~\ref{eq:Helmholtz}}} & \multirow{2}{*}{\makecell{QCPINN}} &  \multirow{2}{*}{\makecell{Angle-Cascade}} & \multirow{2}{*}{771} & $u$ & $6.69\pm1.51$  &  \multirow{2}{*}{$4.52\pm1.86$} & \multirow{2}{*}{\rev{$0.17\pm0.01$}} &  \multirow{2}{*}{\rev{$420.0\pm0.0$}} \\
    
    & & & & $f$ & $2.86\pm0.54$ & \\
    
    \cline{2-9} \noalign{\vskip 1ex}
    
    & \multirow{2}{*}{PINN}&\multicolumn{1}{c}{\multirow{2}{*}{Model-2}}& \multirow{2}{*}{2751}  & $u$  & $12.12\pm7.17$& \multirow{2}{*}{$5.04\pm3.12$} &  \multirow{2}{*}{\rev{$0.002\pm0.01$}} & \multirow{2}{*}{\rev{$340.0\pm0.0$}} \\
    
    & & & & $f$ &  $3.23\pm1.19$ & \\

    \cline{1-9} \noalign{\vskip 1ex}
    
    %%%%%%%%%%%%%%%%%%%%%%%%%%%%%%%%%%%%%%%%% Cavity %%%%%%%%%%%%%%%%%%%%%%%      

    \multirow{6}{*}{\makecell[l]{Cavity\\Eq.~\ref{eq:Cavity}}} &\multirow{3}{*}{\makecell{QCPINN}} & \multirow{3}{*}{Angle-Cascade} & \multirow{3}{*}{923} & $u$ & $ 25.73\pm2.49$  &  \multirow{3}{*}{$0.10\pm0.02$} &  \multirow{3}{*}{\rev{$0.36\pm0.01$}}&  \multirow{3}{*}{\rev{$540.0\pm0.0$}}\\
    &     &  & & $v$ &  $38.00\pm3.76$ &   \\
    &     &  & & $p$ &  $42.65\pm8.69$ &  \\

    \cline{2- 9} \noalign{\vskip 1ex}

    & \multirow{3}{*}{PINN}& \multicolumn{1}{c}{\multirow{3}{*}{Model-1}}&\multirow{3}{*}{8003}  & $u$ & $21.61\pm6.68$& \multirow{3}{*}{$0.06\pm0.02$}  &  \multirow{3}{*}{\rev{$0.01\pm0.0$}}&  \multirow{3}{*}{\rev{$440.0\pm0.0$}}\\
    &     &  & & $v$ & $34.24\pm2.76$& \\
    &     &  & & $p$ & $34.82\pm5.25$& \\
    
    %%%%%%%%%%%%%%%%%%%%%%%%%%%%%%%%%%%%%%%%% Wave %%%%%%%%%%%%%%%%%%%%%%%      

    \cline{1-9} \noalign{\vskip 1ex}
    
    \multirow{2}{*}{\makecell[l]{Wave\\Eq.~\ref{eq:1DWave}}}& \multirow{1}{*}{QCPINN}&  Angle-Cross-mesh& 796& $u$& $12.44\pm4.42$& $0.09\pm0.03$& \rev{$0.38\pm0.08$} & \rev{$469.0\pm0.0$}\\

    \cline{2-9} \noalign{\vskip 1ex} 

    &\multirow{1}{*}{{\makecell{PINN}}}& \multirow{1}{*}{Model-2} &\multirow{1}{*}{2751}  & $u$ & $11.13\pm1.78$& \multirow{1}{*}{$0.05\pm0.01$} & \rev{$0.02\pm0.02$} & \rev{$351\pm0.00$} \\
    
    \cline{1-9} \noalign{\vskip 1ex}

    \multirow{4}{*}{\makecell[l]{Klein-Gordon\\Eq.~\ref{eq:Klein-Gordon}}}& \multirow{2}{*}{QCPINN}& \multirow{2}{*}{Angle-Cascade}& \multirow{2}{*}{771}& $u$& $4.24\pm1.88$&  \multirow{2}{*}{$0.27\pm0.10$} & \multirow{2}{*}{\rev{$0.25\pm0.03$}}& \multirow{2}{*}{\rev{$447.0\pm0.0$}}\\
    &  &  && $f$ & $1.72\pm0.50$ & \\

    \cline{2-9} \noalign{\vskip 1ex}
    
    &\multirow{2}{*}{{\makecell{PINN}}}& \multirow{2}{*}{{Model-2}}& \multirow{2}{*}{2751}  & $u$ & $4.48\pm1.72$& \multirow{2}{*}{$0.34\pm0.13$} & \multirow{2}{*}{\rev{$0.02\pm0.02$}}& \multirow{2}{*}{\rev{$350.0\pm0.0$}}\\
    &  &  && $f$ & $1.79\pm0.47$ &  \\
    \cline{1-9} \noalign{\vskip 1ex}
    
    %%%%%%%%%%%%%%%%%%%%%%%%%%%%%%%%%%%%%%%%% Diffusion %%%%%%%%%%%%%%%%%%%%%%%      

    \multirow{4}{*}{\makecell[l]{Convection-diffusion\\Eq.~\ref{eq:Convection-diffusion}}}& \multirow{2}{*}{QCPINN}& \multirow{2}{*}{Angle-Cross-mesh} & \multirow{2}{*}{821}  & $u$ & $4.54\pm2.17$ &  \multirow{2}{*}{$0.001\pm0.001$} & \multirow{2}{*}{\rev{$0.42\pm0.11$}}& \multirow{2}{*}{\rev{$468.0\pm0.0$}}\\
    &  &  && $f$ & $2.63\pm1.61$ & \\

    \cline{2-9} \noalign{\vskip 1ex}

    &\multirow{2}{*}{{\makecell{PINN}}}&  \multirow{2}{*}{Model-1}&\multirow{2}{*}{7901}  & $u$ & $12.62\pm6.05$& \multirow{2}{*}{$0.02\pm0.02$} & \multirow{2}{*}{\rev{$0.01\pm0.0$}}& \multirow{2}{*}{\rev{$348.0\pm0.0$}}\\
    &  &  && $f$ & $5.57\pm3.05$ &  \\
    \bottomrule
\end{tabular}
\end{table}

%---------------------------------------------------------------------------------------------------------%
\subsection{Helmholtz Equation}
\label{sec:results_dv_helmholtz}
%---------------------------------------------------------------------------------------------------------%

\begin{figure}[t]
    \centering
    \includegraphics[width=0.40\textwidth]{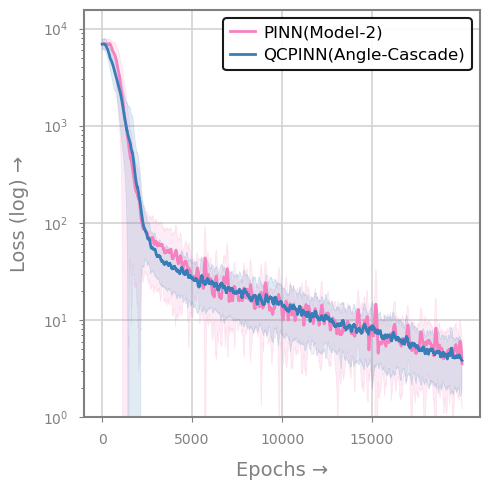}
    \caption{
    \rev{
    Comparison of the loss history of the best models of DV-Circuit QCPINN (Angle-Cascade) with PINN (Model-2) for solving the Helmholtz equation. 
    For loss history of additional models, see Fig.~\ref{fig:loss_history_helmholtz_all} in the appendix.
    }
    }
    \label{fig:loss_history_helmholtz}
\end{figure}

\begin{figure}[t]
\centering
    \includegraphics[width=1\columnwidth, trim={0.20cm 0.20cm 0.20cm 0.20cm}, clip]{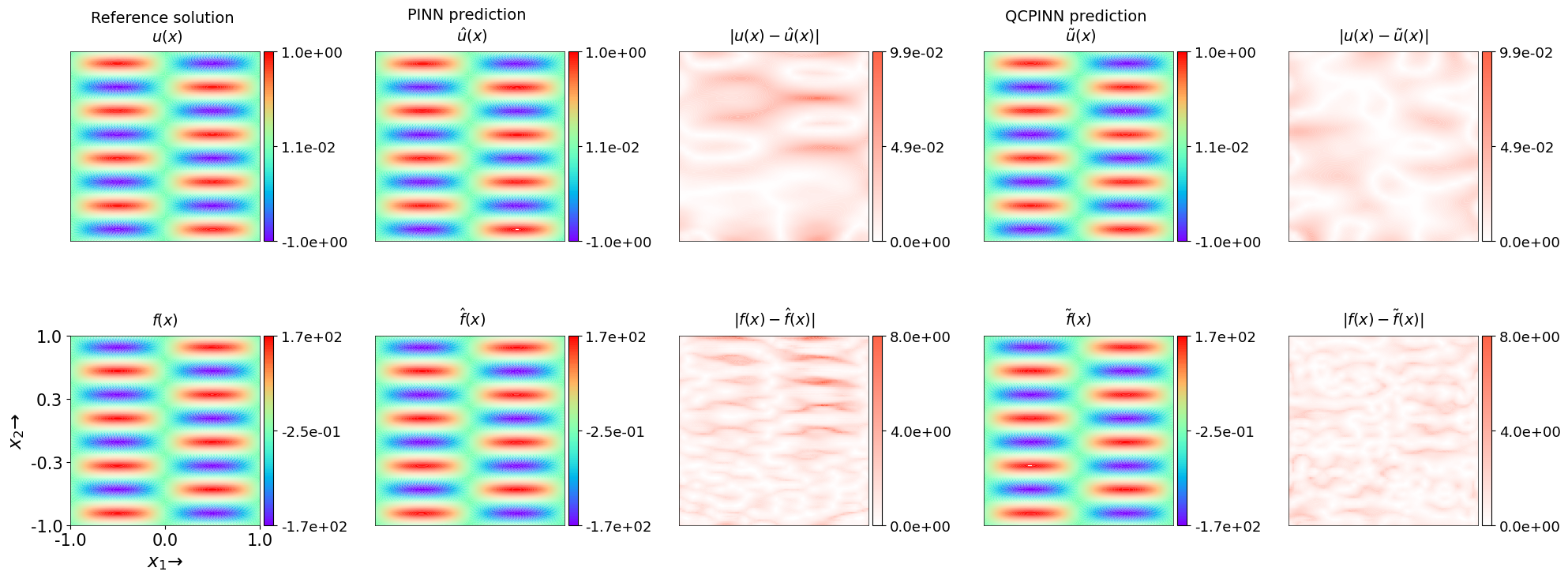}  
    \caption{
    Comparison between the reference solution and model predictions for the Helmholtz equation. This includes the best results from our DV-Circuit QCPINN (Angle-Cascade) alongside PINN (Model-2). The first row shows the PDE solution $u$, while the second row shows the force field $f$. 
    }
\label{fig:helmholtz_prediction}
\end{figure}

Table~\ref{tab:best-results} presents the performance of the best-performing DV-Circuit QCPINN for solving the Helmholtz equation~\ref{eq:Helmholtz} and comparing the results with the best-performing PINN. 
\rev{
The QCPINN (Angle-Cascade) with 771 parameters outperforms the PINN (Model-2) with 2751 parameters, achieving a parameter reduction of $\sim$72\%. This results in a $\sim$45\% lower relative $L_2$ error for the solution field $u$ (6.69\% vs 12.12\%) and 11\% lower error for the force term $f$ (2.86\% vs 3.23\%) at the expense of time and memory. 
}

\rev{
Fig.~\ref{fig:loss_history_helmholtz} compares the training loss history of the best-performing models. The curve shows the mean loss over ten independent runs, with shaded regions indicating the standard deviation. Both models exhibit stable convergence, with the QCPINN achieving slightly lower final loss and reduced variance, indicating more consistent training performance across runs despite using significantly fewer trainable parameters.
}

Fig.~\ref{fig:helmholtz_prediction} presents the reference and predicted solutions, along with the corresponding absolute error distributions. Both QCPINN and PINN models accurately capture the overall solution pattern with visible deviations near the domain boundaries and regions with sharp local gradients.

%---------------------------------------------------------------------------------------------------------%
\subsection{Time-dependent 2D Lid-driven Cavity Equation}
\label{sec:results_dv_cavity}
%---------------------------------------------------------------------------------------------------------%

\begin{figure}[t]
    \centering
    \includegraphics[width=0.40\textwidth]{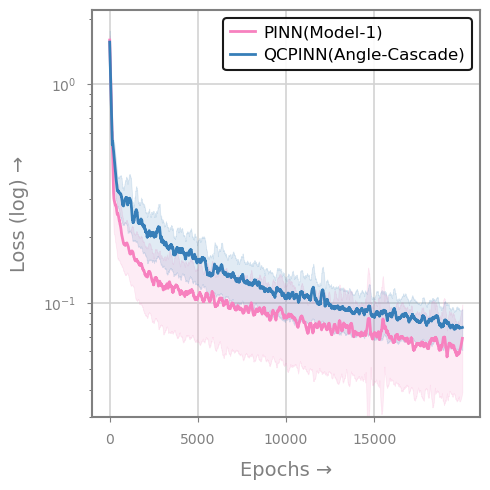}\\
    \caption{ 
    \rev{
    Training loss history for the Cavity problem. The plots compare the convergence behavior of the QCPINN (Angle-Cascade) model with the best-performing PINN (Model-1). 
    }
    }
\label{fig:loss_history_cavity}
\end{figure}

% use 1.0 for IOP
\begin{figure}[t]
\centering        
\ifIOP
    \includegraphics[width=1\columnwidth, trim={0cm 0.0cm 0cm 0.0cm}, clip]{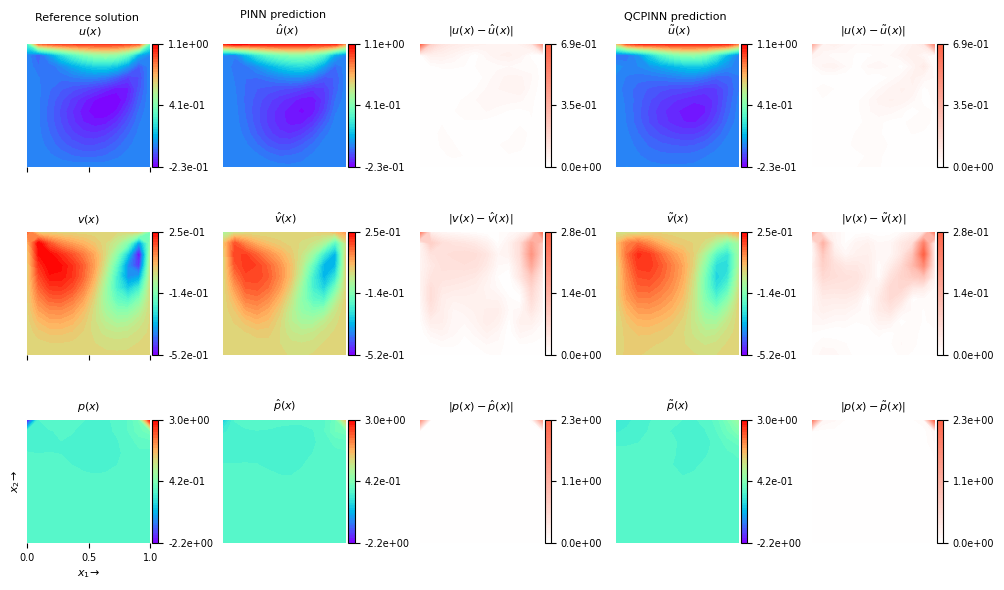} 
\fi
\ifARXIV
    \includegraphics[width=.80\columnwidth, trim={0cm 0.0cm 0cm 0.0cm}, clip]{figures/prediction_cavity.png} 
\fi
\caption{  
Comparison of the reference FEM solution and the model predictions for the lid-driven Cavity problem: the PINN Model-1, and QCPINN (Angle-Cascade). The first and second rows show the velocity fields ($u$ and $v$), while the third row shows the pressure field ($p$). 
}
\label{fig:cavity_prediction}
\end{figure}

\rev{
Table~\ref{tab:best-results} shows that the lid-driven Cavity problem presents a balanced trade-off between parameter efficiency and predictive accuracy. The QCPINN (Angle-Cascade) achieves an $\sim$88\% reduction in trainable parameters compared to PINN Model-1 (923 vs 8003), demonstrating substantial model compactness. Although this reduction is accompanied by modest increases in the relative errors for the velocity components $u$, $v$, and $p$ ($\sim$19\%, $\sim$11\%, and $\sim$23\%, respectively).
}

Fig.~\ref{fig:loss_history_cavity} compares the training loss convergence. Both models achieve stable convergence with minimal oscillations, indicating well-behaved optimization. The QCPINN reaches a slightly higher final loss value ($0.10$ vs $0.06$) as shown in Table~\ref{tab:best-results} but follows a smoother convergence trajectory, suggesting stable learning dynamics even with significantly fewer parameters.

Fig.~\ref{fig:cavity_prediction} visualizes the predicted velocity and pressure fields along with their absolute error maps. Both models reproduce the dominant flow structure, including the central primary vortex and secondary corner eddies, indicating that they effectively capture the overall flow dynamics. However, significant differences emerge in the pressure field. The largest pressure errors for both models occur near the upper moving lid and corner regions, where sharp velocity gradients and strong shear layers lead to steep pressure variations. These regions are inherently challenging for PINN-based models, as the physics-informed loss involves higher-order derivatives that amplify numerical sensitivity in zones with discontinuous boundary forcing. 

%---------------------------------------------------------------------------------------------------------%
\subsection{Additional PDEs}
%---------------------------------------------------------------------------------------------------------%

\begin{figure}[t]
\centering

\begin{tabular}{ccc}
    \includegraphics[width=0.33\columnwidth, trim={0cm .10cm 0cm 0cm}, clip]{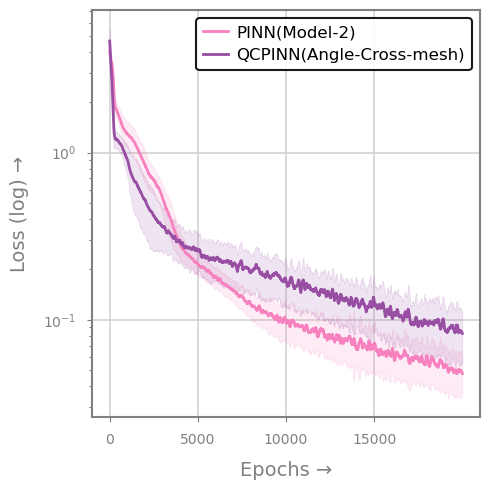} &
    \includegraphics[width=0.33\columnwidth, trim={0cm .10cm 0cm 0cm}, clip]{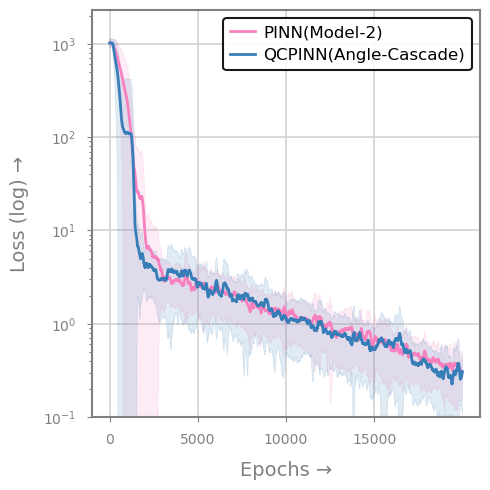} &
    \includegraphics[width=0.33\columnwidth, trim={0cm .10cm 0cm 0cm}, clip]{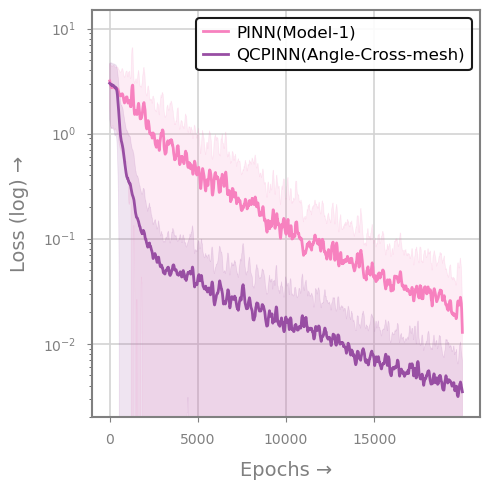} \\
    \footnotesize{(a) Wave} & \footnotesize{(b) Klein-Gordon} & \footnotesize{(c) Convection-diffusion}
\end{tabular}

\caption{
\rev{
Training loss history for additional PDEs. Each plot compares the convergence behavior of the best QCPINN model against the best-performing PINN model.
}
}

\label{fig:more_cases_loss_history}
\end{figure}

\begin{figure}[t]
\centering
\begin{tabular}{c}
    \includegraphics[width=1\columnwidth, trim={0.2cm 0.3cm 0.2cm 0.2cm}, clip]{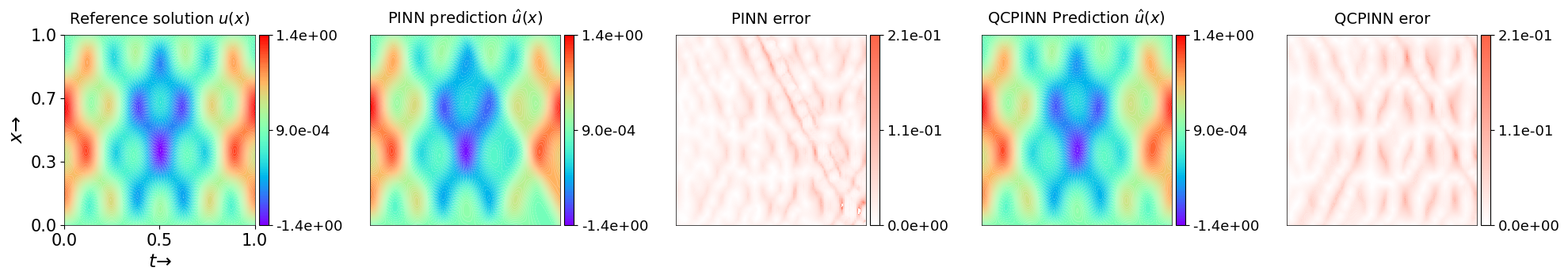}  \\   
    \footnotesize{(a) Wave} \\ 
    \includegraphics[width=1\columnwidth, trim={0.2cm 0.3cm 0.2cm 0.0cm}, clip]{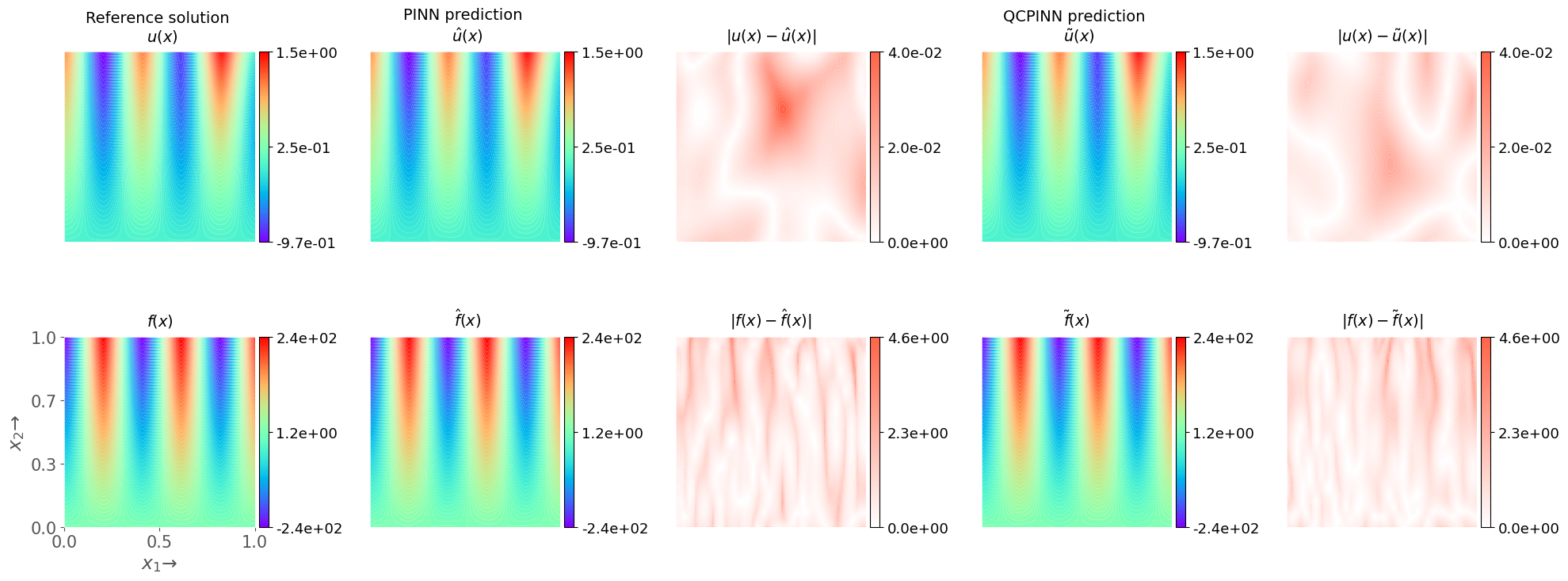}   \\  
    \footnotesize{(b) Klein-Gordon} \\
    \includegraphics[width=1\columnwidth, trim={0.2cm 0.3cm 0.2cm 0.0cm}, clip]{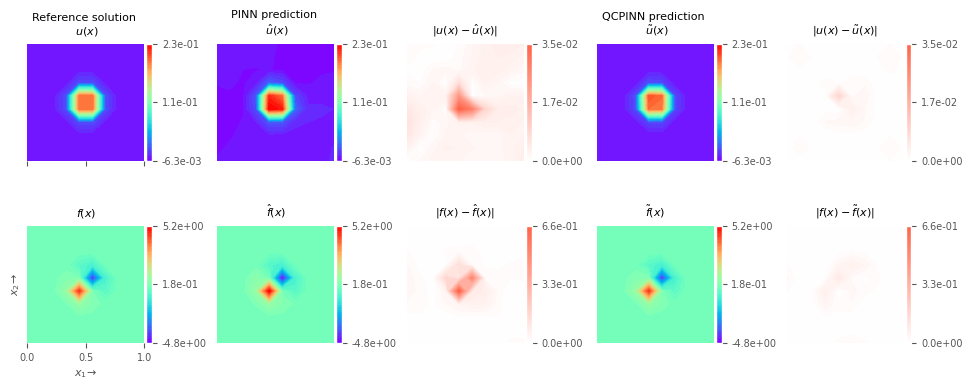} \\
    \footnotesize{(c) Convection-diffusion} \\
\end{tabular}
   
\caption{
Comparison of the reference solution and the model predictions for additional PDEs. 
For each equation, we present results from both the best-performing PINN  and DV-Circuit QCPINN  models. For each equation, the top row displays the velocity field ($u$), while the bottom row (if it exists) shows the force field ($f$). 
}

\label{fig:more_cases_results}
\end{figure}

\rev{
Results from all three additional PDEs (Wave, Klein-Gordon, and Convection-diffusion) consistently show that Angle embedding significantly outperforms Amplitude embedding across all examined cases. 
Additionally, the Cascade and Cross-mesh topologies demonstrated superior performance compared to the other topologies tested. Building on these findings, we conducted further tests on these PDEs using the most effective configurations. The results are presented in Table~\ref{tab:best-results}, while additional results can be found in Table~\ref{tab:additional_cases} in the appendix.
}

\rev{
 For the Wave equation, the Angle-Cross-mesh QCPINN attains a $\sim$71.1\% reduction in parameters (796 vs. 2751) with a slightly higher, $\sim$11.8\% increase, in relative $L_2$ error for $u$ (12.44\% vs. 11.13\%). 
}

\rev{
For the Klein-Gordon equation, the Angle-Cascade QCPINN yields $\sim$72.0\%, parameter savings (771 vs. 2751) while slightly improving accuracy: the mean $L_2$ error for $u$ decreases from 4.48\% to 4.24\%, $\sim$5.4\% relative improvement, and the error for $f$ decreases marginally, $\sim$3.9\%. These differences are small and lie within the reported standard deviations, indicating comparable performance with fewer parameters.
}

\rev{
In Convection-diffusion equation, the Angle-Cascade QCPINN demonstrates both large parameter efficiency (821 vs.\ 7901 parameters, $\sim$89.6\% reduction) and substantially better accuracy: the mean $L_2$ error for $u$ falls from 12.62\% to 4.54\% ($\sim$64.0\% relative improvement), and the error for $f$ decreases from 5.57\% to 2.63\% ($\sim$52.8\% relative improvement).}
\rev{As shown in Table~\ref{tab:additional_cases}, all QCPINN models perform well on this problem. This case shows a clear advantage of the QCPINN architecture in capturing sharp localized features typical of convection-dominated flows.
}

Fig.~\ref{fig:more_cases_loss_history} compares the training loss histories of these models across the additional PDEs. For the Wave equation, Fig.~\ref{fig:more_cases_loss_history}(a) shows that the QCPINN achieves faster initial convergence, reaching lower loss values within the first 4,000 epochs before the baseline PINN (Model-2) begins to converge more rapidly in the later stages.

For the Klein-Gordon equation, Fig.~\ref{fig:more_cases_loss_history}(b) demonstrates that both models follow nearly similar convergence trajectories.
In contrast, for the Convection-diffusion equation, Fig.~\ref{fig:more_cases_loss_history}(c) highlights a clear advantage for the QCPINN, which exhibits a steadily decreasing loss throughout training and attains a final loss approximately an order of magnitude lower than the classical model. This result aligns with Table~\ref{tab:best-results}, where the QCPINN achieved lower relative $L_2$ errors, confirming its better capability in handling convection-dominated dynamics with sharp localized gradients.

%---------------------------------------------------------------------------------------------------------%
\rev{
\subsection{Memory and Time Consideration}
%
%---------------------------------------------------------------------------------------------------------%
%
Across all experiments, the QCPINN models demonstrate a significant increase in computational time per iteration, exhibiting a 15 to 40-fold increase compared to PINN approaches. This discrepancy highlights the computational overhead associated with simulating quantum circuits on classical hardware. 
The memory consumption remains relatively consistent between quantum and classical approaches, ranging from 420 to 540 MB for QCPINN models, compared to 340 to 440 MB for conventional PINN models. This observation suggests that the hybrid architecture does not impose additional memory requirements.
While the computational overhead is a characteristic of the simulator, we anticipate that substantial advantages will become evident as quantum hardware becomes more accessible and practical for implementation. 
}

%---------------------------------------------------------------------------------------------------------%
\rev{
\section{Discussion}
}
%---------------------------------------------------------------------------------------------------------%

\noindent
\rev{
\textbf{Circuit Topology and Entanglement in QCPINN.}
Both the Cross-mesh and Cascade topologies share key structural properties that contribute to their good performance in the QCPINN models. Each employs parameterized controlled-rotation (CRX or CRZ) gates that introduce continuously tunable entanglement between qubits, supporting smoother gradient propagation and mitigating barren-plateau effects during hybrid training (as in Table~\ref{tab:results_helmholtz}, \ref{tab:results_cavity}, and \ref{tab:additional_cases}, and Fig.~\ref{fig:loss_history_helmholtz_all}, \ref{fig:loss_history_cavity_all}, and \ref{fig:more_cases_loss_history_all}). While the Cascade design distributes entanglement efficiently through a ring topology at shallow depth, the Cross-mesh circuit extends this advantage to an all-to-all connectivity pattern. In contrast, the Alternate and Layered circuits rely on fixed CNOT gates, generating discrete, local entangling operations that offer limited tunability and are more prone to optimization plateaus, as reflected in their slower convergence and higher residual losses for the Wave equation (as in Table~\ref{tab:additional_cases} and Fig.~\ref{fig:more_cases_loss_history_all}(a)).
}

\rev{Overall, the comparison indicates that hybrid QCPINN architectures benefit most from hardware-efficient ansätze with continuous entanglement control, such as the Angle-Cascade and Angle-Cross-mesh circuits. These configurations offer a favorable balance between expressivity, trainability, and hardware feasibility, making them highly suitable for near-term quantum solvers of PDEs.}

\noindent
\rev{\textbf{Angle vs. Amplitude Embedding}.
Results consistently demonstrate that Angle embedding achieves better performance compared to Amplitude embedding across all tested configurations (as in Table~\ref{tab:results_helmholtz} and \ref{tab:results_cavity}). 
This advantage arises from both theoretical and practical considerations. 
First, Amplitude embedding represents the data as quantum state amplitudes that must satisfy the normalization constraint $\sum_i |a_i|^2 = 1$, coupling all input features globally and introducing nonlinear dependencies that complicate optimization. 
Second, it scales poorly with input dimensionality, requiring $2^n$ amplitudes to encode $n$ features, which often require zero-padding (or, in our case, we can also use a classical preprocessor with $2^n$ outputs, which increases expressivity by introducing new trainable parameters but does not yield better results, probably due to global coupling through normalization constraint and degraded gradient flow during backpropagation). 
}

\rev{
In contrast, Angle embedding directly maps each classical feature to a rotation angle on a single qubit through parameterized gates such as $RX(\theta)$, allowing local and differentiable embeddings with straightforward gradient computation via automatic differentiation. 
Hence, it provides hardware-efficient, normalization-free feature mapping in which each input independently controls one qubit rotation, yielding stable gradients, better generalization, and faster convergence. 
Empirically, our angle-encoded circuits (e.g., Angle-Cascade) achieved 45–64\% lower relative $L_2$ errors than their amplitude-encoded counterparts across the Helmholtz and Cavity problems (as in Table~\ref{tab:results_helmholtz} and ~\ref{tab:results_cavity}), confirming that feature-to-rotation mappings are better suited for such hybrid QCPINN architectures.
}

\noindent
\rev{
\textbf{Feasibility and Noise Analysis.}
While all experiments in this study were performed on noise-free simulators, practical implementations on near-term quantum processors will inevitably be affected by shot noise from finite measurements, gate infidelity, and decoherence (T1/T2 relaxation). Nevertheless, the feasibility of implementing the proposed QCPINN models on near-term quantum hardware depends on qubit or qumode requirements, circuit depth, entanglement cost, sampling demands, and expected resilience to device noise. 
}

\rev{
For the DV-Circuit QCPINN, the most effective configurations were the Angle-Cascade and Angle-Cross-mesh topologies. Both were implemented with five qubits and a single quantum layer ($L = 1$), as summarized in section~\ref{sec:implementation}. The circuit depth and entangling gate counts vary by topology, as detailed in Table~\ref{tab:topology_comparison}:
\begin{itemize}
    \item The Angle--Cascade circuit follows a ring topology with moderate expressivity and hardware efficiency. Its circuit depth scales as $(n+2)L=7$, using about five entangling gates and roughly 15 trainable parameters. This shallow depth and limited entanglement make it well-suited for current NISQ devices, which typically tolerate 10-20 two-qubit gates before decoherence significantly impacts fidelity \cite{bharti2022noisy}.
    \item The Angle--Cross-mesh circuit adopts all-to-all qubit connectivity. Its depth scales as $(n^{2} - n + 4)L = 24$ for $n=5$, with approximately 20 entangling gates and 45 trainable parameters. Despite being deeper, this layout remained tractable.
\end{itemize}
}

\rev{
For hardware-based training, gradient estimation generally requires $2000$--$5000$ shots per iteration to maintain sampling error below 1\% in observable expectations \cite{cerezo2021variational}. 
Dominant hardware noise sources include gate infidelity, readout errors, and $T_{1}$/$T_{2}$ relaxation. The state-of-the-art superconducting quantum processors achieve single-qubit gate fidelities of 99.9\%~\cite{li2023error,hyyppa2024reducing} and two-qubit fidelities around 99.3-99.7\%~\cite{ding2023high,li2024realization}. Thus, a depth-24 circuit retains effective end-to-end fidelity exceeding 90\% before readout correction.
}

\noindent
\rev{
\textbf{Hyperparameter Sensitivity of CV-Circuit QCPINN Models.}
Although the CV-Circuit QCPINN models did not achieve the same performance levels as their DV-based counterparts, these results remain significant for understanding the current feasibility limits of CV quantum neural networks in physics-informed learning. The observed training failures primarily arose from vanishing gradients imposed by stability constraints, compounded by the intrinsic challenges of PINN optimization, namely, high-order differential requirements and a highly non-convex residual loss landscape. These factors intensified the barren plateau effect, leading to persistent residual loss oscillations with minimal improvement over thousands of iterations (Fig.~\ref{fig:cv_loss_plots}). In other words, the conflict between maintaining valid quantum states (favoring small parameter initialization) and achieving sufficient gradient magnitudes restricted trainability. Among all hyperparameters, the cutoff dimension, learning rate, and initialization scale of active gates (squeezing and displacement amplitudes) exerted the greatest influence on whether training could escape these plateaus.}

\rev{Importantly, the fact that the measurement scheme and nonlinearity choice (as in Table~\ref{tab:qcpinn_config}) had minimal impact on performance suggests that the bottleneck lies not in CV-Circuit expressivity, which remains theoretically superior due to their infinite-dimensional Hilbert spaces, but in the optimization landscape dominated by barren plateaus and restricted gradient flow. This insight directs future research toward developing optimization strategies specifically designed to mitigate barren plateaus in continuous-variable systems (e.g., natural-gradient descent, adaptive parameter-shift rules, layer-wise training, or gradient-free evolutionary methods) rather than further modifying circuit architectures.}

\noindent \textbf{\rev{Limitations and Future Work.}}
Although parameter efficiency was significantly improved, overall solution accuracy remained comparable to baseline PINNs, especially for Cavity and Wave equations, where there is still room for enhancement. 
Additionally, the highly non-convex PINN loss landscape still required careful weighting for the loss terms, which were determined through extensive empirical studies rather than exploring quantum circuit advantage to address this challenge.
Contrary to theoretical expectations, CV-Circuit implementations exhibited training instabilities and underperformed relative to DV-Circuit variants for the cases considered in this study.
\rev{Looking ahead, the promise of quantum-enhanced PINNs in practical applications is evident. Future work should include validation of QCPINN models on quantum hardware to evaluate their resilience to noise and sampling errors-- both critical for real-world applications.
Additionally, developing advanced quantum circuit designs tailored to PDEs and optimizing training stability, especially for CV methods, will further advance the integration of quantum computing and machine learning for solving complex problems across diverse fields.
}

%---------------------------------------------------------------------------------------------------------%
\section{Conclusion} 
\label{sec:conclusion}
%---------------------------------------------------------------------------------------------------------%

We found that QCPINN models achieve comparable solution accuracy to baseline PINNs while requiring a fraction of trainable parameters, as little as 10-30\% of those needed by their classical counterparts under about the same settings.
The experimental evidence supports our hypothesis that hybrid quantum-classical approaches can maintain solution quality while dramatically improving parameter efficiency.
Our systematic assessment of various QCPINN configurations across multiple PDEs consistently revealed that DV-Circuit implementations with Angle embedding and either Cascade or Cross-mesh topologies offer the best performance.
Angle embedding is superior to Amplitude embedding due to its simplicity and gradient stability, as it maps data to individual qubit rotations without normalization constraints, allowing for stable backpropagation. 
The Cascade topology balances expressivity and efficiency, providing enough entanglement to capture complex PDE relationships while keeping circuit depth manageable and avoiding barren plateaus. 
The Cross-mesh topology, on the other hand, leverages all-to-all qubit connectivity to enhance the circuit’s expressivity and its ability to represent globally correlated solution features across the spatial domain of PDEs.  While this design increases the number of entangling gates and circuit depth relative to the Cascade layout, our experiments show that the added expressivity improves generalization and convergence stability without excessive parameter growth.
For the Helmholtz, Klein-Gordon, and Convection-diffusion equations, our QCPINN achieved a significant reduction of around 4–64\% in relative $L_2$ error across various fields compared to baseline PINN models. In contrast, for the Cavity and Wave equations, the $L_2$ error was slightly higher across various fields compared to the baseline PINN models. 
CV-Circuits exhibit persistent training instabilities as standard gradient-based optimization methods, such as Adam, face significant challenges with CV-based quantum circuit models due to their sensitivity to parameter initialization and gradient dynamics.
The findings in this study can be used as a practical guideline for designing hybrid quantum–classical neural architectures to solve PDEs with improved parameter efficiency and stable convergence.
The open-source implementation further enhances reproducibility and enables community engagement. 
 
\section*{Acknowledgment}
We thank the National Center for High-Performance Computing of Turkiye (UHeM) for providing computing resources under grant number 5010662021. 

% keywords can be removed

%Bibliography
\bibliography{sections/references}
\bibliographystyle{unsrt}

\clearpage
\appendix
\section{Loss Function Design for PDEs}
\label{app:pde_difinition}

\subsection{Helmholtz Equation}
The Helmholtz equation represents a fundamental PDE in mathematical physics, arising as the time-independent form of the wave equation. We consider a two-dimensional Helmholtz equation of the form:
\begin{equation}
    \begin{split}
    \Delta u(x,y) + k^2u(x,y) &= f(x,y) \qquad (x,y) \in \Omega   \\ 
    u(x,y) &= h(x,y) \qquad (x,y) \in \Gamma_0
        \label{eq:Helmholtz} 
    \end{split}
\end{equation}

\noindent where $\Delta$ is the Laplacian operator defined as $\Delta u = \frac{\partial^2 u}{\partial x^2} + \frac{\partial^2 u}{\partial y^2}$,  $k$ is the wavenumber (related to the frequency of oscillation), $f(x,y)$ is the forcing term (source function), $h(x,y)$ specifies the Dirichlet boundary conditions.
In this study, we choose an exact solution of the form:
\begin{align*}
    u(x,y) = \sin(a_1\pi x)\sin(a_2\pi y) 
\end{align*}

\noindent on the square domain $\Omega = [-1,1]\times[-1,1]$ with Wavenumber $k = 1$, first mode number $a_1 = 1$, and the  second mode number $a_2 = 4$. This choice of solution leads to a corresponding source term:
\begin{align*}
    f(x,y) = u(x,y)[k^2 - (a_1\pi)^2 - (a_2\pi)^2] \notag
\end{align*}

\noindent  To solve this problem using QCPINN, we construct a loss function that incorporates both the physical governing equation and the boundary conditions:

% TODO
\begin{align*}
\mathcal{L}(\theta) &= \underset{\theta}{\min} \big( \lambda_1\|\mathcal{L}_{\text{phy}}(\theta)\|_{\Omega} + \lambda_2 \|\mathcal{L}_{\text{bc}}(\theta)\|_{\Gamma_0} \big) \\
&= \underset{\theta}{\min} \big( \lambda_1 \text{MSE}(\| \underbrace{u_{\theta_{xx}}(x,y) + u_{\theta_{yy}}(x,y) + \alpha u_{\theta}(x,y)}_{\text{PDE residual}}\|_{\Omega}) + \text{MSE}(\lambda_2 \| \underbrace{u(x,y) - u_{\theta}(x,y)}_{\text{Boundary condition}}\|_{\Gamma_0}) \big) \notag
    % \label{eq:Helmholtz_loss}
\end{align*}

\noindent where $\alpha = (a_1\pi)^2 + (a_2\pi)^2$ combines the mode numbers into a single parameter. The chosen loss weights are $\lambda_1 = 1.0$  and $\lambda_2 = 10.0$.

\subsection{Time-dependent 2D lid-driven cavity Problem}
The lid-driven cavity flow is a classical CFD problem and a fundamental test case for many numerical methods, where flow is governed by the unsteady, incompressible Navier-Stokes equations.  The momentum equation, continuity equation, initial condition, no-slip boundary on walls, and moving lid condition are defined, respectively, as :
\begin{equation}
    \begin{split}
    \rho \left(\frac{\partial \mathbf{u}}{\partial t} + \mathbf{u} \cdot \nabla \mathbf{u}\right) &= - \nabla p + \mu \nabla^2 \mathbf{u} \label{eq:Cavity}\\
    \nabla \cdot \mathbf{u} &= 0  \\
    \mathbf{u}(0 , \mathbf{x}) &= 0 \ \\
    \mathbf{u}(t , \mathbf{x}_0) &= 0 \qquad \mathbf{x} \in \Gamma_0  \\
    \mathbf{u}(t , \mathbf{x}_l) &= 1  \qquad \mathbf{x} \in \Gamma_1 
    \end{split}
\end{equation}

\noindent The computational domain is $\Omega = (0, 1) \times (0, 1)$, 
the spatial discretization, $(N_x, N_y) = (100, 100)$ is uniform grid, with temporal domain $t \in [0, 10]$ seconds and $\Delta t = 0.01s$ with density $\rho = 1056$ kg/m³, 
viscosity $\mu = 1/\mathrm{Re} = 0.01$ kg/(m·s), where Re is the Reynolds number.
$\Gamma_1$ is the top boundary (moving lid) with tangential velocity $U = 1$ m/s, and $\Gamma_0$ is the remaining three sides with no-slip condition ($\mathbf{u} = 0$) where $ \mathbf{u}=(u,v)$.

\noindent For validation purposes, we compare the neural network approximation with results obtained using the finite volume method. The PINN approach employs a composite loss function:
\begin{align*}
    \mathcal{L}(\theta) &=  \underset{\theta}{\min}\big[
    \lambda_1 \|\underbrace{\mathcal{L}_{\text{phy}}(\theta)}_{\text{PDE residual}}\|_{\Omega} 
    + \lambda_2 \|\underbrace{\mathcal{L}_{\text{up}(\theta)} + \mathcal{L}_{\text{bc}_1}(\theta)}_{\text{Boundary conditions}}\|_{\Gamma_1 \cup \Gamma_0} + \lambda_3 \|\underbrace{\mathcal{L}_{\text{u0}}(\theta)}_{\text{Initial conditions}}\|_{\Omega}  \big]
\end{align*}

\noindent where, the physics-informed loss component, $\mathcal{L}_{\mathrm{phy}}(\theta)$ consists of three terms:
\begin{align*}
    \mathcal{L}_{\mathrm{phy}}(\theta) = \mathcal{L}_{r_u} + \mathcal{L}_{r_v} + \mathcal{L}_{r_c}
\end{align*}

\noindent such that:
\begin{align*}
    \mathcal{L}_{r_u}(\theta) &= \text{MSE} \left[(u_{{\theta}_t} + u_{\theta} u_{{\theta}_x} + v_{\theta} u_{{\theta}_y}) + \frac{1.0}{\rho} p_{{\theta}_x} - \mu (u_{{\theta}_{xx}} + u_{{\theta}_{yy}})\right] \\
    \mathcal{L}_{r_v}(\theta) &= \text{MSE} \left[(v_{{\theta}_t} + u_{\theta} v_{{\theta}_x} + v_{\theta} v_{{\theta}_y}) + \frac{1.0}{\rho} p_{{\theta}_y} - \mu (v_{{\theta}_{xx}} + v_{{\theta}_{yy}})\right] \\
    \mathcal{L}_{r_c}(\theta) &= \text{MSE} \left[u_{{\theta}_x} + v_{{\theta}_y}\right] 
\end{align*}

\noindent The boundary and initial conditions are enforced through the following:
\begin{align*}
    \mathcal{L}_{\text{up}} &= \text{MSE}\left[(1.0 - \hat{u}) + \hat{v}\right] \\
    \mathcal{L}_{\text{bc}_{1}} &= \mathcal{L}_{\text{bottom, right, left}} = \text{MSE}\left[\hat{u} + \hat{v}\right] \\
    \mathcal{L}_{\text{u}_{0}} &= \text{MSE}\left[\hat{u} + \hat{v} + \hat{p}\right] 
\end{align*}

\noindent where, $\mathcal{L}_{\mathrm{up}}$ is the moving lid, $\mathcal{L}_{\mathrm{bc1}}$ is the no-slip walls, $\mathcal{L}_{\mathrm{u_0}}$ is the initial conditions. The loss weights are empirically chosen as $\lambda_1 = 0.1$, $\lambda_2 = 2.0$, $\lambda_2 = 2.0$, $\lambda_3 = 4.0$ for $ \mathcal{L}_{\mathrm{phy}}$,  $\mathcal{L}_{\mathrm{up}}$ , $\mathcal{L}_{\mathrm{bc1}}$ and $\mathcal{L}_{\mathrm{u_0}}$ respectively.

\subsection{1D Wave Equation}
The wave equation is a fundamental second-order hyperbolic partial differential equation that models various physical phenomena. In its one-dimensional form, it describes the evolution of a disturbance along a single spatial dimension over time. The general form of the time-dependent 1D wave equation is:
\begin{equation}
    \begin{split}
       u_{tt}(t,x) - c^2u_{xx}(t,x) &= 0  \\ 
        u(t,x_0)  &= f_1(t,x) \qquad (t,x)   \text{ on } \quad  \Gamma_0  \\
        u(t,x_1)  &= f_2(t,x) \qquad (t,x) \text{ on } \quad \Gamma_1  \\
        u(0,x)  &= g(t,x)  \qquad (t,x)  \in \quad \Omega  \\
        u_t(0,x) &= h(t,x) \qquad (t,x)  \in \quad  \partial \Omega  
\label{eq:1DWave}
    \end{split}
\end{equation}
 
\noindent where $u(t,x)$ represents the wave amplitude at position $x$ and time $t$, and $c$ is the wave speed characterizing the medium's properties. The subscripts denote partial derivatives: $u_{tt}$ is the second time derivative, and $u_{xx}$ is the second spatial derivative.

\noindent  For our numerical investigation, we consider a specific case with the following parameters: $c=2$, $a=0.5$,$f_1=f_2=0$. The exact solution is chosen as:
\begin{align*}
u(t,x) = \sin(\pi x) \cos(c\pi t) + 0.5 \sin(2c\pi x) \cos(4c\pi t)
\end{align*}

\noindent  This solution represents a superposition of two standing waves with different spatial and temporal frequencies. The problem is defined on the unit square domain $(t,x) \in [0,1] \times [0,1]$, leading to the specific boundary value problem:
\begin{align*}
        u_{tt}(t,x) -4 u_{xx}(t,x) &= 0 \qquad(t,x) \in \Omega =[0,1] \times [0,1] \\ 
        u(t,0) = u(t,1) &= 0 \\
        u(0,x)  &= \sin(\pi x) + 0.5 \sin(4\pi x)  \quad x \in [0,1] \\
        u_t(0,x) &= 0 
\end{align*}

\noindent To solve this problem using the proposed hybrid QCPINN, we construct a loss function that incorporates the physical constraints, boundary conditions, and initial conditions:
\begin{align*}
\mathcal{L}(\theta) &= \underset{\theta}{\min} \big[ \lambda_1\|\mathcal{L}_{\text{phy}}(\theta)\|_{\Omega}  +   \lambda_2 \|\mathcal{L}_{\text{bc}}(\theta)\|_{\Gamma_1}   +  \lambda_3\|\mathcal{L}_{\text{ic}}(\theta)\|_{\Gamma_0} \big] \\
&= \underset{\theta}{\min} \big[ \lambda_1\text{MSE}( \|\underbrace{u_{\theta_{tt}}(t,x) -4u_{\theta_{xx}}(t,x)}_{\text{PDE residual}}\|_{\Omega})  \\
&+   \lambda_2\text{MSE}( \|\underbrace{u_\theta(t,0) + u_\theta(t,1) + u_\theta(0,x) - \sin(\pi x) - 0.5 \sin(4\pi x) }_{\text{Boundary/initial conditions}}\|_{{\Gamma_0}\cup {\Gamma_1}}) \\
&+  \lambda_3\text{MSE}( \|\underbrace{u_{\theta_{t}}(0,x)}_{\text{Initial velocity}}\|_{\partial \Omega}) \big]
    % \label{eq:1DWave_Loss1}
\end{align*}

\noindent The loss weights are empirically chosen as $\lambda_1 = 0.1$, $\lambda_2 = 10.0$ and $\lambda_3 = 0.1$.
%%%%%%%%%%%%%%%%%%%%%%%%%%%%%%%%%%%%%%%%%%%%%%%%%%%%%%%%%%%%%%%%%%%%%%%%%%%%%%%%%%%%%%
\subsection{Klein-Gordon Equation}
The Klein-Gordon equation is a significant second-order hyperbolic PDE that emerges in numerous theoretical physics and applied mathematics areas. The equation represents a natural relativistic extension of the Schrödinger equation.
In this study, we consider the one-dimensional nonlinear Klein-Gordon equation of the form:
\begin{equation}
    \begin{split}
        u_{tt} - \alpha u_{xx} + \beta u + \gamma u^k &= f(t , x) \qquad(t ,x) \in \Omega\\ 
        u(t,x) &= g_1(t,x) \qquad (t,x) \text{ on } \Gamma_0 \\
        u_t(t,x) &= g_2(t,x)\qquad (t,x) \text{ on } \Gamma_1 \\
        u(0,x) &= h(t,x)  \qquad (t,x) \in \partial \Omega \times [0,T] 
\label{eq:Klein-Gordon}
    \end{split}
\end{equation}

\noindent where $\alpha$ is the wave speed coefficient, $\beta$ is the linear term coefficient, $\gamma$ is the nonlinear term coefficient, and $k$ is the nonlinearity power.
For our numerical study, we set $\alpha = 1$, $\beta = 0$, $\gamma = 1$ and $k = 3$. We choose an exact solution of the form:
\begin{align*}
    u(t,x) = x\cos(5\pi t) + (tx)^3
\end{align*}
This solution combines oscillatory behavior with polynomial growth.
The complete boundary value problem on the unit square domain becomes:
\begin{align*}
        u_{tt}(t,x)- u_{xx}(t,x) + u^3(t,x) &=  x(-25\pi^{2}\cos(5\pi t) - 6t^{3}) \\
        &+ 6t x^{3} + x^{3}(\cos(5\pi t) + t^{3}x^{2})^{3} \quad(t ,x) \in \Omega =[0,1] \times [0,1]\\ 
        u(t,0) &= 0\\
        u(t,1) &= \cos(5\pi t) + t^3  \\
        u(0,x) &= x\\
        u_t(0,x) &= 0
\end{align*}

\noindent To solve this nonlinear PDE using the proposed QCPINN, we construct a composite loss function incorporating the physical constraints, boundary conditions, and initial conditions:
\begin{equation*}
    \begin{split}
\mathcal{L}(\theta) &= \underset{\theta}{\min} \big[ \lambda_1 \|\mathcal{L}_{\text{phy}}(\theta)\|_{\Omega} + \lambda_2 \|\mathcal{L}_{\text{bc}}(\theta)\|_{\Gamma_1} + \lambda_3\|\mathcal{L}_{\text{ic}}(\theta)\|_{\Gamma_0} \big]\\
&= \underset{\theta}{\min} \big[ \lambda_1 \text{MSE}(\|\underbrace{u_{\theta_{tt}}(t,x) - u_{\theta_{xx}}(t,x) + u^3(t,x) -f(t , x)}_{\text{PDE residual}}\|)_{\Omega}\\
&+ \lambda_2 \text{MSE}(\|\underbrace{u(t,0) + u(t,1)-\cos(5\pi t) + t^3 + u(0,x)-x}_{\text{Boundary/initial conditions}}\|)_{\Gamma_1} \\
&+ \lambda_3 \text{MSE}(\|\underbrace{u_{\theta_{t}}(0,x)}_{\text{Initial velocity}}\|_{\partial \Omega}) \big]\\
    \end{split} 
    \label{eq:Klein-Gordon_loss}
\end{equation*}

\noindent where $f(t,x) = x(-25\pi^{2}\cos(5\pi t) - 6t^{3}) + 6t x^{3} + x^{3}(\cos(5\pi t) + t^{3}x^{2})^{3}$. The loss weights are chosen empirically to balance the different components as $\lambda_1 = 1.0$, $\lambda_2 = 10.0$,  and $\lambda_3 = 1.0$.

%%%%%%%%%%%%%%%%%%%%%%%%%%%%%%%%%%%%%%%%%%%%%
\subsection{Convection-diffusion Equation}
The problem focuses on solving the 2D convection-diffusion equation, a partial differential equation that models the transport of a quantity under convection (bulk movement) and diffusion (spreading due to gradients). The specific form considered here includes a viscous equation of the form:
\begin{equation}
\begin{split}
        u_{t} + c_1 u_{x} + c_2 u_{y} - D \Delta u(t,\mathbf{x}) &= f(t,\mathbf{x}) \qquad (t, \mathbf{x}) \in \Omega \\
        u(t, \mathbf{x_0}) &= g_0(t, \mathbf{x_0}) \qquad (t, \mathbf{x_0}) \in \Gamma_0 \\
        u(t, \mathbf{x_1}) &= g_1(t, \mathbf{x_1}) \qquad (t, \mathbf{x_1}) \in \Gamma_1 \\
        u(0, \mathbf{x}) &= h(0 , \mathbf{x}) \qquad \mathbf{x} \in \partial \Omega \times [0,T] 
\label{eq:Convection-diffusion}
\end{split}
\end{equation}

\noindent where $c_1$ is the convection velocity in the $x$ direction, $c_2$ is the convection velocity in the $y$ direction, $D$ is the diffusion coefficient, and $\Delta = \frac{\partial^2}{\partial x^2} + \frac{\partial^2}{\partial y^2}$ is the Laplacian operator. For our numerical investigation, we choose an exact solution describing a Gaussian pulse:
\begin{equation*}
    u(t, x, y) = \exp(-100((x-0.5)^2 + (y-0.5)^2)) \exp(-t)
\end{equation*}

\noindent  The corresponding initial condition is:
\begin{equation*}
    h(0,x,y) = \exp(-100((x-0.5)^2 + (y-0.5)^2))
\end{equation*}

\noindent We choose $c_1 = c_2 = 1.0$, and $D = 0.01$.
The complete initial-boundary value problem on the unit cube domain becomes:
\begin{equation*}
\begin{split}
        u_{t}(t, x, y) &+ u_{x}(t, x, y) + u_{y}(t, x, y) - 0.01\left( u_{xx}(t, x, y) + u_{yy}(t, x, y) \right) = f(t, x, y) \\
        u(t, 0, y) &= \exp(-25 - 100(y-0.5)^2 - t),  \\
        u(t, 1, y) &= \exp(-25 - 100(y-0.5)^2 - t), \\
        u(0, x, y) &= h(x, y) 
\end{split}
\end{equation*}

\noindent where $(t, x, y) \in \Omega = [0,1] \times [0,1] \times [0,1] $, $f(t,x,y) = exp(-100((x-0.5)^2 + (y-0.5)^2) - t)[3 - 200(x-0.5) - 200(y-0.5) - 400((x-0.5)^2 + (y-0.5)^2)]$
To solve this problem using the proposed QCPINN model, we formulate a loss function that incorporates the physical governing equation, boundary conditions, and initial conditions:
\begin{equation*}
\begin{split}
\mathcal{L}(\theta) &= \underset{\theta}{\text{min}} \big( \lambda_1 \|\mathcal{L}_{\text{phy}}(\theta)\|_{\Omega} + \lambda_2 \|\mathcal{L}_{\text{bc}}(\theta)\|_{\Gamma_1} + \lambda_3 \|\mathcal{L}_{\text{ic}}(\theta)\|_{\Gamma_0} \big) \\
&= \underset{\theta}{\text{min}} \big[ \lambda_1 \text{min}(\|\underbrace{u_{\theta_{t}} + u_{\theta_{x}} + u_{\theta_{y}} - 0.01(u_{\theta_{xx}} + u_{\theta_{yy}})}_{\text{PDE residual}}-f\|_{\Omega}) \\
&+ \lambda_2 \text{min}(\|\underbrace{u(t, 0, y) - \exp(-25 - 100(y-0.5)^2 - t)}_{\text{Boundary conditions}}\|_{\Gamma_0} \\
&+ \|\underbrace{u(t, 1, y) - \exp(-25 - 100(y-0.5)^2 - t)}_{\text{Boundary conditions}}\|_{\Gamma_1}) \\
&+ \lambda_3 \text{min}(\|\underbrace{u_{\theta}(0, x, y) - h(x, y)}_{\text{Initial conditions}}\|_{\Omega_0}) \big]
\end{split}
\label{eq:Convection-diffusion_loss}
\end{equation*}

\noindent  The loss weights are chosen to balance the different components$\lambda_1 = 1.0$ , $\lambda_2 = 10.0$, and $\lambda_3 = 10.0$.

% ----------------------------------------------------------

\section{CV-Circuit QCPINN}
\label{app:cv-circuit-qcpinn}

\subsection{Algorithm}
\label{app:cv-circuit-qcpinn-algo}

\begin{algorithm}[ht]
\ifIOP
    \footnotesize
    \caption{
    \rev{
    \footnotesize
    CV-Circuit QCPINN algorithm supports three nonlinear operations: Kerr, cubic phase, and cross-Kerr with flexible measurement schemes: quadrature or number basis. See Section~\ref{sec:cv-training} for training details and the source code~\cite{afrah2025qcpinnGithub} implementation (using PennyLane framework with Strawberry Fields Fock backend). 
    }
    }
\fi
\ifARXIV
    \small
    \captionsetup{font=small}
    \caption{CV-Circuit QNN}
\fi

\begin{algorithmic}[1]
\Require Qumodes: $\text{num\_qumodes}$, Layers: $\text{num\_layers}$, Device: $d$, Cutoff: $c$, Input: Tensor $x$
\Ensure Output: Tensor $y$
\State \textbf{Initialize Parameters:}
\State $\psi_1, \psi_2, \phi_1, \phi_2, \phi_{\text{s}}, \phi_{\text{d}} \sim \mathcal{N}(0, 0.01\pi)$ (interferometer), $\alpha, r \sim \mathcal{N}(0, 0.001)$, $\Phi \sim \mathcal{N}(0, 0.001)$ (nonlinearity)
\State \textbf{Quantum Device:} Initialize $d$ with $\text{num\_qumodes}$, cutoff $c$.
\Procedure{QuantumCircuit}{$\text{input}$}
    \For{$i = 1$ to $\text{num\_qumodes}$}
        \State Encode Inputs with Displacements $D$: $D(\text{input}[i], 0)$ \Comment{only real amplitudes}
    \EndFor
    \For{$l = 1$ to $\text{num\_layers}$}
        \State Interferometer($\psi_1[l]$ , $\phi_1[l]$)
        \State Squeezing($r[l], \phi_{\text{s}}[l]$) \Comment{Phase-free: $\phi_{\text{s}}[l] = 0.0$}
        \State Interferometer($\psi_2[l]$ , $\phi_2[l]$)
        \State Displacement($\alpha[l], \phi_{\text{d}}[l]$) \Comment{Phase-free: $\phi_{\text{d}}[l] = 0.0$}
        \State Nonlinearity $\Phi$ \Comment{Options: Kerr, Cubic Phase, Cross-Kerr}
    \EndFor
    \State Measure state for each wire \Comment{Measurement options: $\langle \hat{q}_i \rangle$ (Quadrature) or $\langle \hat{n}_i \rangle$ (Number)}
    \State \Return Measurements
\EndProcedure
\Procedure{Interferometer}{$\psi$ , $\phi_{\text{bs}}[l]$}
    \For{$l = 1$ to $\text{num\_qumodes}$}
        \For{$(q_1, q_2)$ in Pairs($\text{num\_qumodes}$)}
            \If{$(l + k) \% 2 \neq 1$}
                \State Beamsplitter $B(\psi[n], \phi_{\text{bs}}[n])$ \Comment{Phase-free: $\phi_{\text{bs}}[n] = 0.0$}
            \EndIf
        \EndFor
    \EndFor
    \For{$i = 1$ to $\text{num\_qumodes} - 1$}
        \State Rotation $R_Z(\psi[i])$
    \EndFor
\EndProcedure
\Procedure{ForwardPass}{Input $x$}
        \State $y \gets \text{QuantumCircuit}(x)$
    \State \Return y
\EndProcedure
\end{algorithmic}  
\label{alg:cv-qcpinn}
\end{algorithm}

\subsection{Results}
\label{app:cv-circuit-qcpinn-results}

\begin{table}[t]
\centering
\caption{
The performance of CV-Circuit QCPINN models in solving the Helmholtz and lid-driven Cavity equations. 
There are 12 combinations reported in Table~\ref{tab:qcpinn_config}, however, results with the quadrature measurement scheme, Kerr nonlinear function, and phase-free parameterization are reported here. The remaining 11 combinations are not mentioned as the results are more or less similar.
$N_p$: number of parameters, $L_2$: relative $L_2$ error in \%, $\mathcal{L}_{\text{final}}$: final loss.
Training convergence is shown in~\rev{Fig.~\ref{fig:cv_loss_plots}}.
}
\label{tab:results_cv}

\scriptsize

\begin{tabular}{@{}
        l@{\hspace{4.0pt}} 
        c@{\hspace{8.0pt}}
        c@{\hspace{4.0pt}}
        c@{\hspace{8.0pt}}
        c@{\hspace{8.0pt}}
        c@{\hspace{8.0pt}}
        @{}}
        
    \toprule
    
    \textbf{PDE} & $N_p$ & \multicolumn{2}{c}{$L_2$} & $\mathcal{L}_{\text{final}}$ \\
    \midrule
    
    \multirow{2}{*}{Helmholtz} & \multirow{2}{*}{469} & $u$ & $1.00e+02$  & \multirow{2}{*}{5077.79} \\
                             &  & $f$ &   $1.00e+02$ & &  \\
        \cline{1- 6} \noalign{\vskip 1ex}
  \multirow{3}{*}{Cavity} & \multirow{3}{*}{621} & $u$ & $9.345e+01$  & \multirow{3}{*}{0.102} \\
                             &  & $v$ &   $9.979e+01$ & &  \\
                             &  & $p$ &   $8.620e+01$ & &  \\
\bottomrule
\end{tabular}
\end{table}

\begin{figure}[htb]
\centering
\begin{tabular} {
    c@{\hspace{1.0pt}} @{\hspace{1.0pt}}
    c@{\hspace{1.0pt}} @{\hspace{1.0pt}}
}  
\ifIOP
    \includegraphics[width=0.49\columnwidth, trim={0.35cm 0.35cm 0.20cm 0.25cm}, clip]{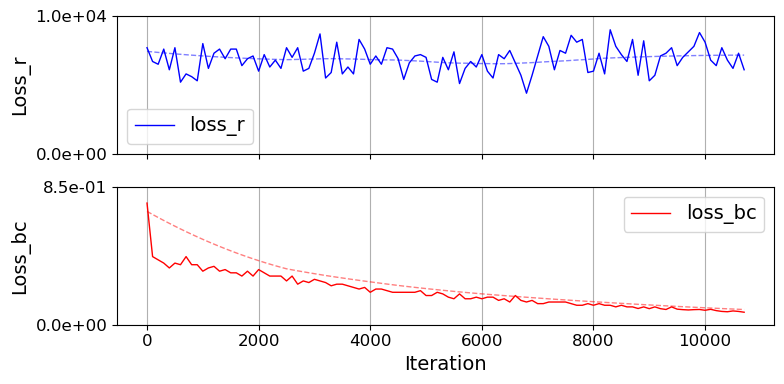} &
    \includegraphics[width=0.49\columnwidth,trim={0.15cm 0.35cm 0.20cm 0.25cm}, clip]{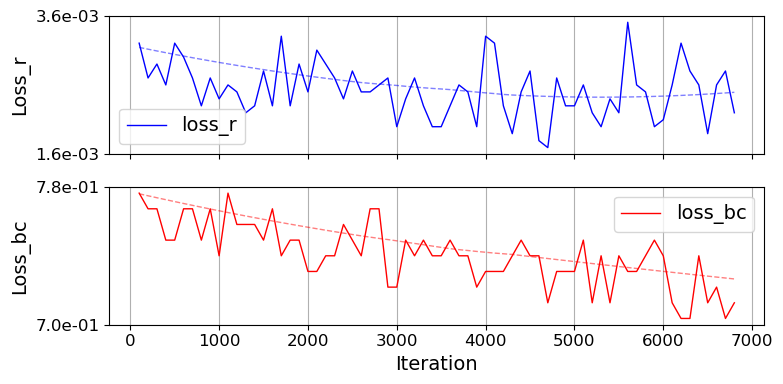} \\
    \footnotesize{(a) Helmholtz}  & \footnotesize{(b) Cavity} \\
\fi
\ifARXIV
    \includegraphics[width=0.39\columnwidth, trim={0.35cm 0.35cm 0.20cm 0.25cm}, clip]{figures/loss_plots_helmholtz.png} &
    \includegraphics[width=0.39\columnwidth,trim={0.15cm 0.35cm 0.20cm 0.25cm}, clip]{figures/loss_plots_Cavity.png} \\
    \footnotesize{(a) Helmholtz}  & \footnotesize{(b) Cavity} \\
\fi
    
\end{tabular}
\caption{The training convergence of the CV-Circuit QCPINN model when solving Helmholtz and lid-driven Cavity. The terms ``loss\_r'' and ``loss\_bc'' refer to residual and boundary losses. The ``loss\_bc'' shows a decreasing trend, indicating improvements in satisfying the boundary conditions; however, the overall decrease is below expectation. The ``loss\_r'' exhibits significant fluctuations and does not display a noticeable reduction, suggesting difficulties in minimizing the residual error of the governing equation.
}

\label{fig:cv_loss_plots}
\end{figure}

\vspace{2mm}\noindent
\textbf{Helmholtz Equation.} 
For the Helmholtz Eq.~\ref{eq:Helmholtz}, we used 469 trainable parameters. The results, in Table~\ref{tab:results_cv}, show a high relative $L_2$ error for both the $u$ and $f$ variables, along with a significant final loss. 
The training convergence pattern in Fig.~\ref{fig:cv_loss_plots}(a) highlights some challenges. For example, the boundary loss, ``loss\_bc'', shows a steady decrease over 10,000 iterations, suggesting that the model is gradually improving in meeting the boundary conditions. However, the residual loss, ``loss\_r'', remains consistently high, with substantial fluctuations throughout the training process. This indicates that the model was struggling to satisfy the differential equation itself.

\vspace{2mm}\noindent
\textbf{Time-dependent 2D Lid-driven Cavity Equation.}
For the more complex lid-driven Cavity Eq.~\ref{eq:Cavity}, the CV-Circuit QCPINN model was implemented with 621 trainable parameters. The result, in Table~\ref{tab:results_cv}, shows that the relative $L_2$ errors for both velocity and pressure variables remain high. However, the final loss value of 0.102 is significantly lower than that observed for the Helmholtz equation. 
Fig.~\ref{fig:cv_loss_plots}(b) shows the training convergence, where the boundary loss exhibits some improvement, although it still experiences considerable fluctuations. The residual loss displays even greater oscillations compared to the Helmholtz case, indicating a greater challenge in accurately capturing the complex fluid dynamics governed by the Navier-Stokes equations.

This behavior is primarily attributed to gradient decay issues intrinsic to our CV-Circuit implementation. The vanishing gradient problem~\cite{glorot2010understanding} manifests more severely due to the additional constraints imposed by quantum operations. Small initialization values, while necessary to prevent divergence in quantum state evolution, significantly diminish gradients during backpropagation.

We examined various techniques, such as Xavier initialization for learnable parameters, gradient clipping, and skip connections, to mitigate the problem of gradient decay, but these attempts proved ineffective. Exploring alternative optimization strategies beyond standard gradient descent may be beneficial in addressing this issue, though this remains beyond the scope of this work.

\clearpage

\section{DV-Circuit QCPINN}
\label{app:dv_circuit_all}

\begin{table}[htb]
\centering
\caption{
Results from all configurations of QCPINN (DV-Circuit) models and PINN models for solving the Helmholtz equation. $N_p$: number of trainable parameters, $L_2$: relative error (in \%), $L_\text{final}$: final loss, T: time per iteration (in sec), M: peak memory consumption (in MB), $u$: reference solution, and $f$ source term.
}
\label{tab:results_helmholtz}
\scriptsize

\begin{tabular}{@{}l
        l@{\hspace{14.0pt}}
        c@{\hspace{4.0pt}}@{\hspace{4.0pt}} 
        c@{\hspace{14.0pt}}
        c@{\hspace{14.0pt}}
        c@{\hspace{14.0pt}}
        c@{\hspace{14.0pt}}
        c@{\hspace{14.0pt}}
        c@{\hspace{14.0pt}}
        @{}}
        \toprule
        
    & \textbf{Embedding}& \textbf{Topology}& $N_p$& \multicolumn{2}{c}{$L_2$}&  $\mathcal{L}_{\text{final}}$ & \rev{T} & \rev{M}\\

    \midrule

    \multirow{18}{*}{\makecell{QCPINN\\(DV-Circuit)}} & \multirow{8}{*}{Amplitude} & \multirow{2}{*}{Alternate} & \multirow{2}{*}{772}  & $u$& $31.98\pm9.51$ &  \multirow{2}{*}{$24.42\pm10.91$} &  \multirow{2}{*}{\rev{$0.26\pm0.00$}} &  \multirow{2}{*}{\rev{$420.0\pm0.0$}} \\
    
    & & & &  $f$ & $7.70\pm1.42$ & \\
    
    \cline{3-9} \noalign{\vskip 1ex}
    
    & & \multirow{2}{*}{Cascade} & \multirow{2}{*}{771}  & $u$& $21.33\pm8.19$ &  \multirow{2}{*}{$35.49\pm29.26$} &  \multirow{2}{*}{\rev{$0.28\pm0.00$}} &  \multirow{2}{*}{\rev{$420.0\pm0.0$}} \\
    
    & & & & $f$ & $8.03\pm2.83$ & \\
    
    \cline{3-9} \noalign{\vskip 1ex}

    &  &  \multirow{2}{*}{Cross-mesh} & \multirow{2}{*}{796} & $u$& $29.39\pm17.84$&  \multirow{2}{*}{$23.29\pm16.08$} &  \multirow{2}{*}{\rev{$0.31\pm0.0$}} &  \multirow{2}{*}{\rev{$440.0\pm0.0$}} \\
    
    &     &  & &  $f$ &  $7.30\pm3.21$ & \\
    
    \cline{3-9} \noalign{\vskip 1ex}

    & &  \multirow{2}{*}{Layered} & \multirow{2}{*}{776} & $u$& $20.95\pm7.46$&  \multirow{2}{*}{$18.16\pm9.18$} &  \multirow{2}{*}{\rev{$0.29\pm0.03$}} &  \multirow{2}{*}{\rev{$420.0\pm0.0$}} \\
     
     &  &  & &  $f$ &  $4.96\pm0.92$ &  \\
    
    \cline{2-9} \noalign{\vskip 1ex}

    &  \multirow{8}{*}{\makecell{Angle}}  & \multirow{2}{*}{Alternate} & \multirow{2}{*}{772}  & $u$& $10.74\pm5.54$ &\multirow{2}{*}{$4.87\pm3.19$} &  \multirow{2}{*}{\rev{$0.18\pm0.0$}} &  \multirow{2}{*}{\rev{$420.0\pm0.0$}} \\
    
    &     &  & &  $f$ & $3.42\pm0.8$ &   \\
    
    \cline{3-9} \noalign{\vskip 1ex}

    &  &  \multirow{2}{*}{Cascade} & \multirow{2}{*}{771} & $u$ & $\mathbf{6.69\pm1.51}$  &  \multirow{2}{*}{$4.52\pm1.86$} &  \multirow{2}{*}{\rev{$0.17\pm0.00$}} &  \multirow{2}{*}{\rev{$420.0\pm0.0$}} \\
    
    &     &  & &  $f$ &  $\mathbf{2.86\pm0.54}$ &   \\
    
    \cline{3-9} \noalign{\vskip 1ex}

    &  &  \multirow{2}{*}{Cross-mesh} & \multirow{2}{*}{796} & $u$& $7.24\pm1.45$ &  \multirow{2}{*}{$3.69\pm1.30$} &  \multirow{2}{*}{\rev{$0.25\pm0.02$}} &  \multirow{2}{*}{\rev{$430.0\pm0.0$}} \\
    
    & &  &  & $f$ &  $3.09\pm0.70$ &  \\
    
    \cline{3-9} \noalign{\vskip 1ex}

    & &  \multirow{2}{*}{Layered} & \multirow{2}{*}{776} & $u$& $18.19\pm10.24$&  \multirow{2}{*}{$11.63\pm8.73$} &  \multirow{2}{*}{\rev{$0.19\pm0.00$}} &  \multirow{2}{*}{\rev{$420.0\pm0.0$}} \\
    
    & & & & $f$ & $4.88\pm1.42$ &  \\
        
    \midrule
        
    \multirow{4}{*}{PINN}& \multicolumn{2}{c}{\multirow{2}{*}{Model-1}}& 
    \multirow{2}{*}{7851}  & $u$  & $21.24\pm12.13$ & \multirow{2}{*}{$16.62\pm13.85$} &  \multirow{2}{*}{\rev{$0.0041\pm0.0$}} &  \multirow{2}{*}{\rev{$340.0\pm0.0$}} \\
    
    & & & & $f$ & $5.59\pm2.56$  & \\
    
    \cline{2-9} \noalign{\vskip 1ex}
    
    & \multicolumn{2}{c}{\multirow{2}{*}{Model-2}}& \multirow{2}{*}{2751}  & $u$  & $12.12\pm7.17 $& \multirow{2}{*}{$5.04\pm3.12 $} &  \multirow{2}{*}{\rev{$0.0027\pm0.0$}} &  \multirow{2}{*}{\rev{$340.0\pm0.0$}} \\
    
    & & & & $f$ &  $3.23\pm1.19$ & \\
\bottomrule
\end{tabular}
\end{table}

\begin{figure}[htb]
\centering
\ifIOP
    \includegraphics[width=0.50\columnwidth, trim={0cm 0cm 0cm 0cm}, clip]{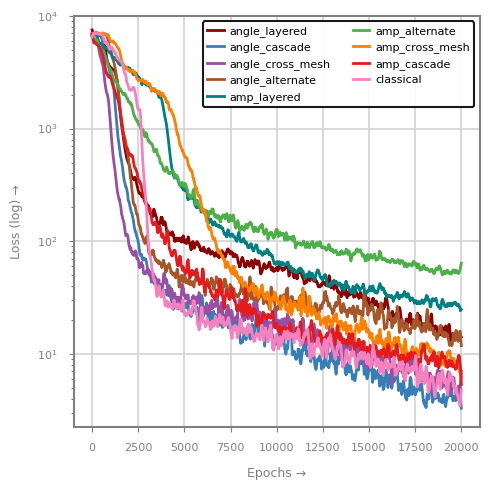}
\fi
\ifARXIV
    \includegraphics[width=0.35\columnwidth, trim={0cm 0cm 0cm 0cm}, clip]{figures/loss_history_helmholtz_all.png}
\fi
\caption{
Training convergence history of all QCPINN (DV-Circuit) models compared to the best classical PINN model when solving the Helmholtz equation.
}
\label{fig:loss_history_helmholtz_all}
\end{figure}

\begin{table}[htb]
\centering
\caption{
Results from all configurations of QCPINN (DV-Circuit) models and PINN models for solving the Cavity problem. $N_p$: number of trainable parameters, $L_2$: relative error (in \%), $L_\text{final}$: final loss, T: time per iteration (in sec), M: peak memory consumption (in MB), $u$ and $v$: velocity, and $p$: pressure.
}
\label{tab:results_cavity}
\scriptsize

\begin{tabular}{@{}l
        c@{\hspace{8.0pt}}
        c@{\hspace{4.0pt}}@{\hspace{4.0pt}}
        c@{\hspace{8.0pt}}
        c@{\hspace{8.0pt}}
        c@{\hspace{8.0pt}}
        c@{\hspace{8.0pt}}
        c@{\hspace{8.0pt}}
        c@{\hspace{8.0pt}}
        @{}}
        
    \toprule
    
    & \textbf{Embedding}& \textbf{Topology}& $N_p$&  & $L_2 $ & $\mathcal{L}_{\text{final}}$ & \rev{T} & \rev{M}\\

    \midrule
  
    \multirow{26}{*}{\makecell{QCPINN}} &  \multirow{11}{*}{\makecell{Amplitude}}    &   \multirow{3}{*}{Alternate} & \multirow{3}{*}{924}  & $u$ & $54.40\pm7.56$ &  \multirow{3}{*}{$0.08\pm0.01$}&  \multirow{3}{*}{\rev{$0.57\pm0.0$}} &  \multirow{3}{*}{\rev{$570.0\pm0.0$}}\\
    &     &  & & $v$ & $83.07\pm1.04$ &   \\
    &     &  & & $p$ & $55.53\pm4.38$ &   \\
    \cline{3- 9} \noalign{\vskip 1ex}

    &    &   \multirow{3}{*}{Cascade} & \multirow{3}{*}{928}  & $u$ & $59.70\pm2.61$ &  \multirow{3}{*}{$0.10\pm0.03$} &  \multirow{3}{*}{\rev{$0.69\pm0.07$}} &  \multirow{3}{*}{\rev{$580.0\pm0.0$}}\\
    &     &  & & $v$ &  $80.15\pm3.38$ &  \\
    &     &  & & $p$ &  $56.44\pm7.54$ &  \\
    \cline{3- 9} \noalign{\vskip 1ex}
    
    &  & \multirow{3}{*}{Cross-mesh} & \multirow{3}{*}{948} & $u$ & $39.95\pm0.00$&  \multirow{3}{*}{$0.07\pm0.00$} &  \multirow{3}{*}{\rev{$0.65\pm0.0$}} &  \multirow{3}{*}{\rev{$620.0\pm0.01$}} \\
    &     &  & & $v$ &  $49.62\pm1.10$ & \\
    &     &  & & $p$ &  $57.13\pm2.00$ & \\
    \cline{3- 9} \noalign{\vskip 1ex}
    
    && \multirow{3}{*}{Layered} & \multirow{3}{*}{928} & $u$ & $59.92\pm3.08$&  \multirow{3}{*}{$0.09\pm0.03$} &  \multirow{3}{*}{\rev{$0.60\pm0.02$}} &  \multirow{3}{*}{\rev{$590.0\pm0.0$}}\\
    &     &  & & $v$ &  $82.67\pm3.33$ &  \\
    &     &  & & $p$ &  $51.92\pm1.99$ &  \\
    
    \cline{2- 9} \noalign{\vskip 1ex}
  
    &  \multirow{11}{*}{\makecell{Angle}}   &  \multirow{3}{*}{Alternate} & \multirow{3}{*}{924}  & $u$ & $26.13\pm2.00$ &\multirow{3}{*}{$0.05\pm0.00$} &  \multirow{3}{*}{\rev{$0.37\pm0.0$}} &  \multirow{3}{*}{\rev{$540.0\pm0.0$}} \\
    &     &  & & $v$ & $43.86\pm9.00$ &    \\
    &     &  & & $p$ & $43.86\pm4.10$ &    \\
    \cline{3- 9} \noalign{\vskip 1ex}

    &  & \multirow{3}{*}{Cascade} & \multirow{3}{*}{923} & $u$ & $ 
    25.73\pm2.49$  &  \multirow{3}{*}{$0.10\pm0.02$} &  \multirow{3}{*}{\rev{$0.36\pm0.0$}} &  \multirow{3}{*}{\rev{$540.0\pm0.0$}} \\
    &     &  & & $v$ &  $38.00\pm3.76$ &   \\
    &     &  & & $p$ &  $
    42.65\pm8.69$ &  \\
    \cline{3- 9} \noalign{\vskip 1ex}

     &  & \multirow{3}{*}{Cross-mesh} & \multirow{3}{*}{948} & $u$ & $27.30\pm4.87$ &  \multirow{3}{*}{$0.08\pm0.03$} &  \multirow{3}{*}{\rev{$0.47\pm0.0$}} &  \multirow{3}{*}{\rev{$590.0 \pm0.0$}}\\ 
    &     &  &  & v &  $39.88\pm4.58$ &  \\
    &     &  &  & p &  $44.99\pm9.24$ &  \\
    \cline{3- 9} \noalign{\vskip 1ex}

    & & \multirow{3}{*}{Layered} & \multirow{3}{*}{928} & $u$ & $59.87\pm11.71$ &  \multirow{3}{*}{$0.1\pm0.17$} &  \multirow{3}{*}{\rev{$4.01\pm0.02$}} &  \multirow{3}{*}{\rev{$560.0\pm0.0$}} \\ 
    &     &  &  & v &  $85.41\pm8.22$ & \\
    &     &  &  & p &  $73.36\pm13.45$ & \\
    \cline{1- 9} \noalign{\vskip 1ex}
    
    \multirow{6}{*}{PINN}& \multicolumn{2}{c}{\multirow{3}{*}{Model-1}}&\multirow{3}{*}{8003}  & $u$ & $\mathbf{21.61\pm6.68}$ & \multirow{3}{*}{$0.06\pm0.02$}  &  \multirow{3}{*}{\rev{$0.01\pm0.0$}} &  \multirow{3}{*}{\rev{$440.0\pm0.0$}} \\
    &     &  & & $v$ & $\mathbf{34.24\pm2.76}$& \\
    &     &  & & $p$ & $\mathbf{34.82\pm5.25}$& \\
    
    \cline{2- 9} \noalign{\vskip 1ex}
    
    &\multicolumn{2}{c}{\multirow{3}{*}{Model-2}}& \multirow{3}{*}{2903}  & $u$ & $23.03\pm2.45$ & \multirow{3}{*}{$0.06\pm0.02$}  &  \multirow{3}{*}{\rev{$0.01\pm0.0$}} &  \multirow{3}{*}{\rev{$435\pm0.0$}}\\
    &     &  & & $v$ & $38.81\pm2.73$ & \\
    &     &  & & $p$ & $71.51\pm4.79$& \\
\bottomrule
\end{tabular}
\end{table}
 
\begin{figure}[htb]
\centering
\ifIOP
    \includegraphics[width=0.50\columnwidth, trim={0cm 0cm 0cm 0cm}, clip]{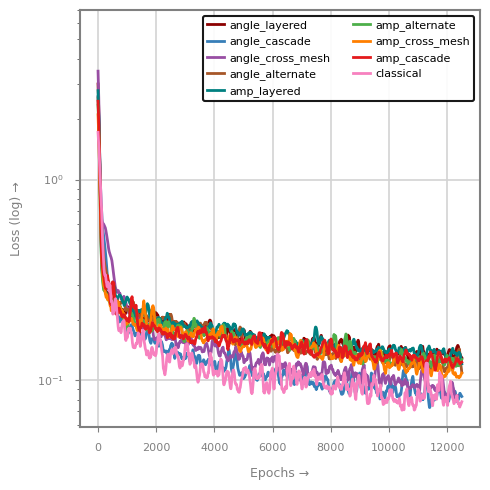}\\
\fi
\ifARXIV
   \includegraphics[width=0.40\columnwidth, trim={0cm 0cm 0cm 0cm}, clip]{figures/loss_history_cavity_all.png}
\fi
\caption{
Training convergence history of all QCPINN (DV-Circuit) models compared to the best classical PINN model when solving the Cavity problem.
}
\label{fig:loss_history_cavity_all}
\end{figure}

\begin{table}[htb]
\centering
\caption{
Results from selected configurations of QCPINN (DV-Circuit) models and two PINN models. $N_p$: number of trainable parameters, $L_2$: relative error (in \%), $L_\text{final}$: final loss, T: time per iteration (in sec), M: peak memory consumption (in MB), $u$: velocity, and $f$: source terms.
}
\label{tab:additional_cases}

\scriptsize

\begin{tabular}{@{}l
        c@{\hspace{12.0pt}}
        lll
        c@{\hspace{8.0pt}}
        c@{\hspace{12.0pt}}
        c@{\hspace{8.0pt}}
        c@{\hspace{8.0pt}}
        c@{\hspace{8.0pt}}
        @{}}
    
    \toprule
        
    \textbf{PDEs}& \multicolumn{2}{c}{\textbf{Method}}& $N_p$& &$L_2$ &$\mathcal{L}_{\text{final}}$ & \rev{T} & \rev{M} \\
    
    \midrule

    %%%%%%%%%%%%%%%%%%%%%%%%%%%%%%%%%%%%%%%%% Wave %%%%%%%%%%%%%%%%%%%%%%%      
    
    \multirow{6}{*}{Wave Eq.~\ref{eq:1DWave}}& \multirow{4}{*}{QCPINN}& Angle-Alternate& 772& $u$& $39.96\pm7.85$& $0.78\pm0.46$& \rev{$0.15\pm0.0$} & \rev{$405.0\pm0.0$}\\
    & & Angle-Layered& 776& $u$& $44.76\pm0.28$& $1.13\pm0.26$& \rev{$0.17\pm0.0$} & \rev{$419.0\pm0.0$}\\
    & & Angle-Cascade& 771& $u$& $16.74\pm5.41$& $0.10\pm0.03$& \rev{$0.25\pm0.06$} & \rev{$446.0\pm0.0$}\\

    % \cline{3-9} \noalign{\vskip 1ex}

    & & Angle-Cross-mesh& 796& $u$& $12.44\pm4.42$& $0.09\pm0.03$& \rev{$0.38\pm0.08$} & \rev{$469.0\pm0.0$} \\

    \cline{2-9} \noalign{\vskip 1ex} 

    &\multirow{2}{*}{{\makecell{PINN}}}& \multirow{1}{*}{Model-1}&\multirow{1}{*}{$7851$}  & $u$ & $20.94\pm1.70$& \multirow{1}{*}{$0.15\pm0.04$}& \rev{$0.02\pm0.03$} & \rev{$353.0\pm0.0$} \\
    
    % \cline{3-9} \noalign{\vskip 1ex}

    &&\multirow{1}{*}{Model-2} &\multirow{1}{*}{2751}  & $u$ & $\mathbf{11.13\pm1.78}$& \multirow{1}{*}{$0.05\pm0.01$} & \rev{$0.02\pm0.02$} & \rev{$351\pm0.00$} \\
    
    \cline{1-9} \noalign{\vskip 1ex}
%%%%%%%%%%%%%%%%%%%%%%%%%%%%%%%%%%%%%%%%% Klein %%%%%%%%%%%%%%%%%%%%%%%      

    \multirow{12}{*}{\makecell{Klein-Gordon\\Eq.~\ref{eq:Klein-Gordon}}}& \multirow{8}{*}{QCPINN}& \multirow{2}{*}{Angle-Alternate}& \multirow{2}{*}{772}& $u$& $15.54\pm5.56$&  \multirow{2}{*}{$0.51\pm0.37$} & \multirow{2}{*}{\rev{$0.16\pm0.0$}} & \multirow{2}{*}{\rev{$406.0\pm0.0$}} \\
    &  &  && $f$ & $1.99\pm0.96$ & \\

    & & \multirow{2}{*}{Angle-Layered}& \multirow{2}{*}{776} & $u$& $14.86\pm4.78$&  \multirow{2}{*}{$0.51\pm0.41$} & \multirow{2}{*}{\rev{$0.17\pm0.0$}} & \multirow{2}{*}{\rev{$409.0\pm0.0$}} \\
    &  &  && $f$ & $2.33\pm1.10$ & \\

    & & \multirow{2}{*}{Angle-Cascade}& \multirow{2}{*}{771} & $u$& $\mathbf{4.24\pm1.88}$&  \multirow{2}{*}{$0.27\pm0.10$} & \multirow{2}{*}{\rev{$0.25\pm0.03$}} & \multirow{2}{*}{\rev{$447.0\pm0.0$}} \\
    &  &  && $f$ & $\mathbf{1.72\pm0.50}$ & \\
    % \cline{3-9} \noalign{\vskip 1ex}

    & & \multirow{2}{*}{Angle-Cross-mesh}& \multirow{2}{*}{796} & $u$& $5.64\pm1.87$&  \multirow{2}{*}{$0.34\pm0.21$} & \multirow{2}{*}{\rev{$0.42\pm0.11$}} & \multirow{2}{*}{\rev{$468.0\pm0.0$}} \\
    &  &  && $f$ & $1.65\pm0.38$ & \\
    \cline{2-9} \noalign{\vskip 1ex}
    
    &\multirow{4}{*}{{\makecell{PINN}}}& \multirow{2}{*}{Model-1} &\multirow{2}{*}{7851}  & $u$ & $5.47\pm3.57$& \multirow{2}{*}{$0.61\pm0.35$} & \multirow{2}{*}{\rev{$0.03\pm0.04$}} & \multirow{2}{*}{\rev{$348.0\pm0.0$}} \\
    &  &  && $f$ & $2.58\pm1.24$ &  \\
    % \cline{3-9} \noalign{\vskip 1ex}

    &&\multirow{2}{*}{{Model-2}}& \multirow{2}{*}{2751}  & $u$ & $4.48\pm1.72$& \multirow{2}{*}{$0.34\pm0.13$} & \multirow{2}{*}{\rev{$0.02\pm0.02$}} & \multirow{2}{*}{\rev{$350.0\pm0.0$}} \\
    &  &  && $f$ & $1.79\pm0.47$ &  \\
    \cline{1-9} \noalign{\vskip 1ex}
%%%%%%%%%%%%%%%%%%%%%%%%%%%%%%%%%%%%%%%%% Diffusion %%%%%%%%%%%%%%%%%%%%%%%    

    \multirow{12}{*}{\makecell{Convection-diffusion\\Eq.~\ref{eq:Convection-diffusion}}}& \multirow{8}{*}{QCPINN}& \multirow{2}{*}{Angle-Alternate} & \multirow{2}{*}{822}  & $u$ & $5.06\pm2.08$ &  \multirow{2}{*}{$0.001\pm0.001$} & \multirow{2}{*}{\rev{$0.13\pm0.0$}} & \multirow{2}{*}{\rev{$403.0\pm0.0$}} \\
    &  &  && $f$ & $3.22\pm0.65$ & \\

    & & \multirow{2}{*}{Angle-Layered} & \multirow{2}{*}{826}  & $u$ & $5.10\pm3.22$ &  \multirow{2}{*}{$0.01\pm0.01$} & \multirow{2}{*}{\rev{$0.15\pm0.0$}} & \multirow{2}{*}{\rev{$410.0\pm0.0$}} \\
    &  &  && $f$ & $3.40\pm1.62$ & \\
    
    & & \multirow{2}{*}{Angle-Cascade} & \multirow{2}{*}{821}  & $u$ & $5.04\pm1.43$ &  \multirow{2}{*}{$0.001\pm0.001$} & \multirow{2}{*}{\rev{$0.27\pm0.06$}} & \multirow{2}{*}{\rev{$445.0\pm0.0$}} \\
    &  &  && $f$ & $3.03\pm1.00$ & \\
    % \cline{3-9} \noalign{\vskip 1ex}

    & & \multirow{2}{*}{Angle-Cross-mesh} & \multirow{2}{*}{846}  & $u$ & $\mathbf{4.54\pm2.17}$ &  \multirow{2}{*}{$0.001\pm0.001$} & \multirow{2}{*}{\rev{$0.42\pm0.11$}} & \multirow{2}{*}{\rev{$468.0\pm0.0$}} \\
    &  &  && $f$ & $\mathbf{2.63\pm1.61}$ & \\
    \cline{2-9} \noalign{\vskip 1ex}

    &\multirow{4}{*}{{\makecell{PINN}}}&  \multirow{2}{*}{Model-1}&\multirow{2}{*}{7901}  & $u$ & $12.62\pm6.05$& \multirow{2}{*}{$0.02\pm0.02$} & \multirow{2}{*}{\rev{$0.01\pm0.0$}} & \multirow{2}{*}{\rev{$348.0\pm0.0$}}\\
    &  &  && $f$ & $5.57\pm3.05$ &  \\
    % \cline{3-9} \noalign{\vskip 1ex}

    &&\multirow{2}{*}{{Model-2}}& \multirow{2}{*}{2801}  & $u$ & $36.80\pm8.79$& \multirow{2}{*}{$0.11\pm0.06$} & \multirow{2}{*}{\rev{$0.01\pm0.0$}} & \multirow{2}{*}{\rev{$346.0\pm0.0$}} \\
    &  &  && $f$ & $21.27\pm8.12$ &  \\
    \bottomrule
\end{tabular}
\end{table}

\begin{figure}[htb]
\centering

\ifIOP
\begin{tabular}{ccc}
    \includegraphics[width=0.33\columnwidth, trim={0cm .10cm 0cm 0cm}, clip]{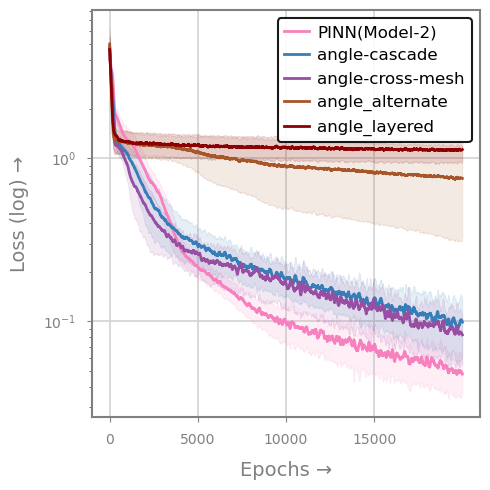} &
    \includegraphics[width=0.33\columnwidth, trim={0cm .10cm 0cm 0cm}, clip]{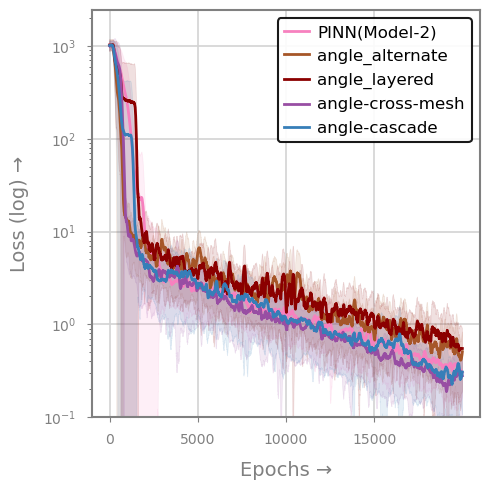} &
    \includegraphics[width=0.33\columnwidth, trim={0cm .10cm 0cm 0cm}, clip]{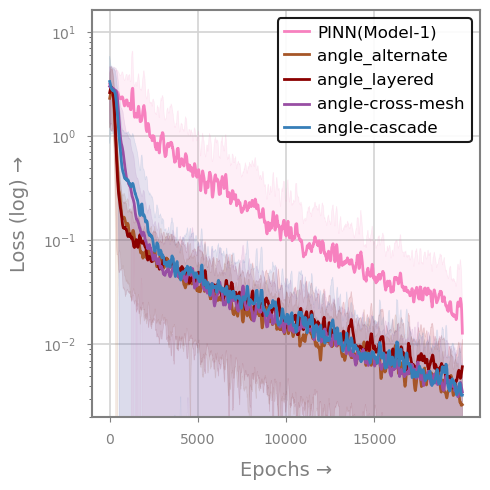} \\
    \footnotesize{(a) Wave} & \footnotesize{(b) Klein-Gordon} & \footnotesize{(c) Convection-diffusion}
\end{tabular}
\fi

\ifARXIV
\begin{tabular}{ccc}    
    \includegraphics[width=0.25\columnwidth, trim={0cm .10cm 0cm 0cm}, clip]{figures/loss_history_wave_all.png} &
    \includegraphics[width=0.25\columnwidth, trim={0cm .10cm 0cm 0cm}, clip]{figures/loss_history_klein_all.png} &
    \includegraphics[width=0.25\columnwidth, trim={0cm .10cm 0cm 0cm}, clip]{figures/loss_history_diffusion_all.png} \\
    \footnotesize{(a) wave} & \footnotesize{(b) Klein-Gordon} & \footnotesize{(c) Convection-diffusion}
\end{tabular}
\fi

\caption{ 
Each plot compares the convergence behavior of the  QCPINN models with angle embedding with the best-performing classical PINN model.
}

\label{fig:more_cases_loss_history_all}
\end{figure}

\end{document}